\documentclass[11pt,a4paper]{article}

\usepackage[T1]{fontenc}
\usepackage[utf8]{inputenc}
\usepackage[left=3.0cm,right=3.0cm,top=3.2cm,bottom=3.2cm]{geometry}
\usepackage[strict]{changepage}
\usepackage{amsmath,amssymb,amsthm,mathrsfs}
\usepackage{booktabs}
\usepackage{xcolor}
\usepackage[numbers,sort&compress]{natbib}
\usepackage{hyperref}
\usepackage{titlesec}
\usepackage{caption}
\usepackage{subcaption}
\usepackage{float}
\usepackage{enumitem}
\usepackage{pgfplots}
\pgfplotsset{compat=1.17}
\usepackage{tikz}
\usetikzlibrary{arrows.meta,positioning,calc}
\usepackage{tcolorbox}
\tcbuselibrary{skins,breakable}
\usepackage{array}
\usepackage{listings}
\definecolor{qblue}{RGB}{0,70,140}        
\definecolor{qorange}{RGB}{210,130,0}
\definecolor{qgreen}{RGB}{0,158,115}
\definecolor{qred}{RGB}{213,94,0}
\definecolor{qpurple}{RGB}{148,103,189}
\definecolor{qgray}{RGB}{100,100,100}
\definecolor{qboxbg}{RGB}{237,244,251}
\definecolor{codebg}{RGB}{248,248,248}

\hypersetup{colorlinks,linkcolor=qblue,citecolor=qblue,urlcolor=qblue}

\titleformat{\section}{\large\bfseries}{\thesection}{0.7em}{}[\vspace{-2pt}]
\titleformat{\subsection}{\normalsize\bfseries}{\thesubsection}{0.6em}{}
\titleformat{\subsubsection}{\normalsize\itshape}{\thesubsubsection}{0.5em}{}
\titlespacing{\section}{0pt}{10pt plus2pt}{4pt plus1pt}
\titlespacing{\subsection}{0pt}{7pt plus1pt}{2pt plus1pt}

\theoremstyle{remark}
\newtheorem{remark}{Remark}

\newtcolorbox{keyresult}[1][]{
  colback=qboxbg, colframe=qblue,
  fonttitle=\bfseries\small, title={#1},
  boxrule=0.5pt, left=6pt, right=6pt, top=4pt, bottom=4pt, breakable,
  arc=2pt}

\pgfplotsset{
  qplot/.style={
    grid=major,
    grid style={line width=0.25pt, draw=qgray!25},
    tick label style={font=\small},
    label style={font=\small},
    title style={font=\small\bfseries, yshift=-2pt},
    tick align=inside,
    line width=0.9pt,
  }
}

\newcommand{\Pket}{|{P}\rangle}    
\newcommand{\Rket}{|{R}\rangle}    
\newcommand{\DeltaE}{\Delta E}

\begin{document}

\begin{center}
\vspace*{8pt}
{\LARGE\bfseries
Classically Forbidden Signatures \\ of Quantum Coherence\\[4pt]
in the Mesoscopic Lipkin--Meshkov--Glick Model}\\[16pt]

{\normalsize\bfseries Stavros Mouslopoulos}\\[4pt]
{\small\itshape
Faculty of Science and Engineering, University of Nottingham Ningbo China \\
Ningbo 315100, China}\\[3pt]
{\small\texttt{Stavros.Mouslopoulos@nottingham.edu.cn}}\\[12pt]
{}
\rule{0.8\linewidth}{0.4pt}
\end{center}


\begin{abstract}
We derive strict quantitative conditions under which a collective quantum
system of $N\sim370$ spins exhibits classically forbidden temporal
correlations in a spinor Bose--Einstein condensate (BEC).
The Lipkin--Meshkov--Glick (LMG) model near its $\mathbb{Z}_2$-breaking
quantum critical point supports a mesoscopic superposition
$\alpha\Pket+\beta\Rket$ of two macroscopic ordered phases
($\Pket$: $m_z\approx+m_*$; $\Rket$: $m_z\approx-m_*$)
at the Goldilocks crossover $N\approx N_c$, where the tunnel splitting
equals $k_BT$ and macroscopic susceptibility $\chi\sim N$ coexists
with finite quantum coherence.
We establish two quantum-discriminating predictions.
\textbf{P4 (Landau--Zener crossover, proposed discriminator):}
quantum tunnelling drives $P_{\rm error}\to0$ exponentially with quench
time, while the classically non-ergodic system remains kinetically frozen
at $P_{\rm error}\to1$ --- a parametrically large, computable separation
that rests on a specific kinetic foil.
\textbf{P5 (Leggett--Garg inequality violation, strictly model-independent):}
$K_3>1$ is forbidden by macrorealism for \emph{all} classical models
satisfying non-invasive measurability.
A five-level Lindblad simulation yields $K_3\approx1.32$ and dephasing
threshold $\gamma_\phi\lesssim0.289$~s$^{-1}$ at the BEC target
($\gamma_\phi=0.05$~s$^{-1}$, $N=370$, $\Gamma/J=0.95$) ---
a margin of $8.8\times$ above the optimal measurement interval,
well within current experimental reach.
The threshold has a precise physical origin in the emergent collective
$\mathbb{Z}_2$ symmetry: exact parity eliminates the dominant dephasing
cross-term and renders the LGI correlator immune to $T_1$ population
mixing without requiring ground-state preparation ($2.35\times$
improvement over the naive mean-field estimate); constructive dynamical
phase alignment of higher odd-parity states at the benchmark parameters
contributes a further $2.47\times$.
All results are reproducible from the provided self-tested Python code.
\end{abstract}

\vspace{4pt}
\noindent\rule{\linewidth}{0.3pt}
\vspace{4pt}

\section{Introduction}
\label{sec:intro}
 
The Lipkin--Meshkov--Glick (LMG) model~\cite{Lipkin1965} is a paradigmatic
exactly-solvable collective spin system, whose full $(N+1)$-level spectrum
is obtained by diagonalising a tridiagonal matrix in the symmetric Dicke
subspace.
It possesses a $T=0$ quantum critical point at $\Gamma_c=J$
(at which the two ordered vacua become exactly degenerate in the
thermodynamic limit), has been studied in
entanglement entropy~\cite{Latorre2005}, finite-size
scaling~\cite{Botet1983}, non-equilibrium Kibble--Zurek
dynamics~\cite{Zurek1996,Defenu2018,Puebla2020,Kopylov2019}, and is directly realised
in spinor Bose--Einstein condensates (BECs)~\cite{Zibold2010,Trenkwalder2016}.
 
The central question of this paper is: \emph{can the LMG model, operating
in an open-system environment at experimentally accessible parameters,
produce signatures of quantum coherence that are classically forbidden ---
and if so, what are the strict quantitative conditions?}
In particular, we ask whether the Leggett--Garg inequality (LGI),
$K_3\leq1$ for any macrorealist model~\cite{LeggettGarg1985,Emary2014},
can be violated --- and by how much, under what dephasing, and via what
physical mechanisms --- in a mesoscopic spinor BEC near the LMG critical
point.
This question matters at two levels.
At the level of condensed matter physics, the LMG model is the simplest
collective quantum phase transition with a binary order parameter and exact
$\mathbb{Z}_2$ symmetry, and establishing quantitative LGI violation
thresholds for it provides a model system against which more complex
proposals --- molecular nanomagnets, polariton condensates, mixed-species
BECs --- can be benchmarked.
At a deeper level, four structural features that make the LGI protocol
work here are not peculiarities of the LMG model: they are properties of
any $\mathbb{Z}_2$-symmetric collective bistable quantum system near its
critical point.
Specifically: (i)~the collective parity symmetry protects the LGI
correlator from $T_1$-type population noise exactly within the
parity-protected ground doublet, requiring no ground-state preparation;
(ii)~the $K_3>1$ prediction uses the standard operational quantum description
of projective measurements via the Lüders instrument, formulated at the
density-matrix level without invoking selective collapse in individual runs;
(iii)~the inequality bound is interpretation-independent,
ruling out all macrorealist models satisfying non-invasive measurability simultaneously;
and (iv)~the violation is robust against known proposals for
non-unitary modifications to quantum mechanics, at a level quantified in
Section~\ref{sec:conclusion}.
All four features are established in the present paper; a companion
paper~\cite{LMGpreprint} applies the same structural framework in a
broader context, building on the quantitative foundation developed here.
 
\textbf{The two macroscopic phases and the mesoscopic window.}
The model supports two macroscopic ordered phases, labelled $\Pket$
(order parameter $m_z\approx+m_*>0$) and $\Rket$ ($m_z\approx-m_*<0$).
Precisely, $\Pket$ and $\Rket$ are spin-coherent states centred at the
mean-field minima $m_z=\pm m_*=\pm\sqrt{1-\Gamma^2/J^2}$; they are
\emph{not} the Dicke basis states $|{+}{+}\cdots{+}\rangle$ and
$|{-}{-}\cdots{-}\rangle$ (approached only as $\Gamma\to0$).
Their overlap is $S=\langle P|R\rangle=(\Gamma/J)^N=(1-m_*^2)^{N/2}$~\cite{Arecchi1972}
(evaluated for the spin-coherent state ansatz centred at $\pm m_*$,
not for exact localised eigenstates of the double well),
exponentially small at the parameters used throughout:
$S\approx5.7\times10^{-9}$ at $\Gamma/J=0.95$, $N=370$, so two-level
approximations treating $|P\rangle$ and $|R\rangle$ as orthogonal are
excellent within the Goldilocks zone.
For finite $N$ the true energy eigenstates are well approximated by
$|E_0\rangle=(|P\rangle+|R\rangle)/\sqrt{2+2S}$ and
$|E_1\rangle=(|P\rangle-|R\rangle)/\sqrt{2-2S}$, separated by the
tunnel splitting $\DeltaE(N)$.
The Goldilocks crossover $N\approx N_c$, where $\DeltaE(N_c)=k_BT$,
is the regime of interest: $N\ll N_c$ gives deep quantum coherence but
small susceptibility; $N\gg N_c$ gives classical thermal dominance.
At $N\approx N_c$ the system simultaneously exhibits macroscopic
susceptibility $\chi\sim N$ and coherent tunnelling between $\Pket$
and $\Rket$, creating the window for classically forbidden signatures.
For the benchmark BEC parameters used throughout ($\Gamma/J=0.95$,
$T=10$~nK), exact diagonalisation gives $N_c\approx370$ and
$\DeltaE/\hbar=1310$~rad/s~$=k_BT/\hbar$.
 
\textbf{The two quantum-discriminating predictions.}
Classical signatures of near-critical LMG dynamics --- critical slowing
down, oscillatory autocorrelation, Kibble--Zurek scaling --- are
reproducible by classical nonlinear bistable systems.
We identify two that are not.
\textbf{P4 (Landau--Zener crossover, proposed discriminator).}
In a spinor BEC, spin thermalisation operates on timescales of
seconds~\cite{Trenkwalder2016}, far exceeding experimentally accessible
quench (sweep) times $\tau_Q\sim\text{ms}$.
The classical Kramers escape time $\tau_K\propto\Gamma_{\rm eff}^{-1}
e^{N\Delta f_0/k_BT}$ --- where $\Delta f_0$ is the mean-field free-energy
barrier per spin --- is exponentially large: at our benchmark parameters
$N\Delta f_0/k_BT\approx13$, giving $\tau_K\sim130$~s, so the classical
system is non-ergodic on all experimental timescales.
It therefore cannot follow the sweep and remains frozen in its initial
well, giving an adiabatic transition error probability $P_{\rm error}\to1$
for all accessible $\tau_Q\ll\tau_K$.
The quantum LMG system, within the coherent window
$\hbar/\DeltaE\ll\tau_Q\ll T_2^{\rm phys}$ (where $T_2^{\rm phys}$ is the
physical eigenstate coherence time defined in Section~\ref{sec:lindblad})
and sweep amplitude
$\DeltaE/(Nm_*)\ll\Delta h\ll|h_{\rm sp}|$ (where $h_{\rm sp}$ is the
spinodal field at which the metastable well disappears,
Section~\ref{sec:predictions}), tunnels coherently and drives
$P_{\rm error}$ exponentially to zero via the Landau--Zener mechanism.
The quantum--classical separation in $P_{\rm error}(\tau_Q)$ is
parametrically large and computable, but rests on a specific kinetic foil
(Model A classical dynamics, Section~\ref{sec:mf}) and is therefore
presented as a proposed discriminator complementing the strictly
model-independent P5.
\textbf{P5 (Leggett--Garg Inequality violation).}
For any macrorealist model satisfying non-invasive measurability,
the three-time correlator $K_3 = C_{12}+C_{23}-C_{13}$ of the
dichotomic-like observable $Q=\operatorname{sgn}(\hat{J}_z)$ satisfies
$K_3\leq1$~\cite{LeggettGarg1985,Emary2014}.
The quantum LMG predicts $K_3$ up to
$Q_{01}^2\times\frac{3}{2}=1.415$ (decoherence-free ceiling,
where $Q_{01}=\langle E_0|Q|E_1\rangle$ is the off-diagonal observable
matrix element between the ground doublet states).
Sakamoto \textit{et al.}~\cite{Sakamoto2024} demonstrated LGI violation
in closed-system LMG dynamics in principle; the present paper establishes
the strict quantitative open-system conditions for the first time.
 
A key result of this paper is that the LGI dephasing threshold has three
analytically distinct levels, each capturing a different layer of the physics.
The \emph{mean-field} level~A ($\gamma_\phi^{(A)}\lesssim0.050$~s$^{-1}$)
uses the mean-field coherent-state estimate $T_2^{\rm coll}$ and places
the BEC target exactly at the threshold ($K_3\approx1.000$).
The \emph{analytic eigenstate} level~B ($\gamma_\phi^{(B)}\lesssim0.117$~s$^{-1}$)
corrects the parity-forbidden cross-term, raising the threshold $2.35\times$
and giving $K_3\approx1.22$ at the BEC target.
The \emph{five-level Lindblad} level~C ($\gamma_\phi^{(C)}\lesssim0.289$~s$^{-1}$)
captures reinforcing contributions from higher odd-parity states, raising
it a further $2.47\times$ and giving $K_3\approx1.32$ at the BEC target.
The $5.8\times$ total improvement (derived in
Remarks~\ref{rem:hierarchy} and~\ref{rem:reconcile}) converts a borderline
mean-field prediction into a robust, experimentally unambiguous result.
The underlying rationale is this: near but not at criticality, the LMG
model is not simply two-level, and the naive two-level (mean-field)
approximation fails badly in the coherent-state basis.
The exact collective $\mathbb{Z}_2$ symmetry, however, eliminates the
dominant failure mode — the dephasing cross-term — when the correct
energy-eigenstate basis is used, rehabilitating a symmetry-protected
quasi-two-level description.
The higher-level corrections are then small, computable, and at the
benchmark parameters net positive, so the full five-level simulation
provides the authoritative threshold with negligible further correction
from levels beyond five.

\textbf{Contributions.}
The argument proceeds in three layers: classical kinetic analysis
(Section~\ref{sec:mf}), Goldilocks crossover (Section~\ref{sec:goldilocks}),
and open-system quantum predictions with full numerical verification
(Sections~\ref{sec:predictions}--\ref{sec:multilevel}).
The specific contributions are:
(a) the exact $Q_{01}^2=0.9436$ matrix element from complete $371$-dimensional
symmetric Dicke sector diagonalisation, which sets the decoherence-free
ceiling $K_3^{\rm max}=1.415$;
(b) the three-level threshold hierarchy (A/B/C) identifying the parity
cross-term elimination ($2.35\times$) and the net multi-level Dicke-doublet
contribution ($2.47\times$) as physically distinct mechanisms,
together raising the BEC target $K_3$ from borderline ($1.000$)
to robust ($1.32$);
(c) a five-level Lindblad simulation validated against ten-level results
to $\sim1\%$, yielding the authoritative dephasing threshold
$\gamma_\phi\lesssim0.289$~s$^{-1}$;
(d) explicit evaluation of both local and collective bath decoherence
scaling, establishing why the collective bath is the relevant model
for BEC global field noise;
(e) the Landau--Zener crossover as a complementary, experimentally
direct protocol for discriminating quantum tunnelling from classical
non-ergodic trapping;
(f) a controlled semiclassical WKB instanton analysis quantifying the
Goldilocks crossover $N_c$ and its approximation error
(severely inaccurate at $\Gamma/J\gtrsim0.98$, where exact diagonalisation
is mandatory); and
(g) complete, self-tested Python code for BEC experimental calibration,
reproducing all figures and thresholds.
 
\section{LMG Hamiltonian and Phase Structure}
\label{sec:ham}

The LMG Hamiltonian with Kac normalisation~\cite{Kac1968}:
\begin{equation}
  \hat{H} = -\frac{J}{2N}\!\left(\sum_{i=1}^N\hat\sigma_i^z\right)^{\!2}
    - \Gamma\sum_{i=1}^N\hat\sigma_i^x
    - h\sum_{i=1}^N\hat\sigma_i^z.
  \label{eq:lmg}
\end{equation}
In terms of collective spin operators
$\hat{J}_z=\frac{1}{2}\sum_i\hat\sigma_i^z$,
$\hat{J}_\pm=\hat{J}_x\pm i\hat{J}_y$:
$\hat{H}=-\frac{2J}{N}\hat{J}_z^2-2\Gamma\hat{J}_x-2h\hat{J}_z$.
BEC mapping~\cite{Zibold2010}: $J\leftrightarrow N(g_{12}-g_{11})$,
$\Gamma\leftrightarrow\hbar\Omega$, $h\leftrightarrow\hbar\delta$.\footnote{%
$g_{11}$ and $g_{12}$ are the intra- and inter-species $s$-wave scattering
interaction strengths respectively; $\Omega$ is the two-photon Rabi coupling
frequency between the two internal hyperfine states; $\delta$ is the
corresponding laser detuning (all in rad\,s$^{-1}$).
The ordered phase ($\Gamma<J$) requires $J>0$, i.e.\ $g_{12}>g_{11}$
(inter-species repulsion exceeds intra-species), which is achievable
via Feshbach resonance tuning~\cite{Trenkwalder2016}.}
Throughout, $\hbar=k_B=1$; all energies and rates are given in
rad\,s$^{-1}$ (setting $\hbar=1$ means rad\,s$^{-1}$ and s$^{-1}$
are dimensionally equivalent).\footnote{Physical times quoted in milliseconds are
obtained by restoring $\hbar$: e.g.\ $T_2^{\rm phys}=(\hbar/\DeltaE)\times(\DeltaE\,T_2^{\rm phys})$,
with $\DeltaE/\hbar=1310$~rad\,s$^{-1}$ giving $\hbar/\DeltaE\approx0.763$~ms.}
At these parameters, $k_BT=1310$~rad/s at $T=10$~nK, and the tunnel splitting
$\DeltaE/\hbar=1310$~rad/s at $N=370$, $\Gamma/J=0.95$ (exact diagonalisation
of the complete $371$-dimensional symmetric Dicke sector matrix).
The intensive order parameter is $m_z\equiv N^{-1}\sum_i\langle\hat\sigma_i^z\rangle$.
Throughout, $m_z\in[-1,1]$ denotes the intensive order parameter
and $m\in\{-N/2,\ldots,N/2\}$ the Dicke eigenvalue of $\hat{J}_z$.

The mean-field variational free energy per spin (exact only in the
thermodynamic mean-field limit; used here as the effective thermodynamic
landscape for the classical kinetic foil) at temperature $T$:
\begin{equation}
  f(m) = \tfrac{J}{2}m^2 - T\ln\!\left[2\cosh\!\tfrac{\sqrt{(Jm+h)^2+\Gamma^2}}{T}\right].
  \label{eq:freeenergy}
\end{equation}
The quantum critical point is at $T=0$, $\Gamma_c=J$~\cite{Sachdev2011}.
In the ordered phase ($\Gamma<J$, $h=0$), two macroscopic vacua emerge at
$m_*=\pm\sqrt{1-\Gamma^2/J^2}$, which we label:
\begin{equation}
\begin{split}
  &\Pket:\; m_z\to +m_*
    \;(\text{spin-coherent state at }+m_*;\text{ approaches }|{+}{+}\cdots{+}\rangle\text{ as }\Gamma\to0), \\
  &\Rket:\; m_z\to -m_*
  \;(\text{spin-coherent state at }-m_*;\text{ approaches }|{-}{-}\cdots{-}\rangle\text{ as }\Gamma\to0).
\end{split}
\label{eq:vacua}
\end{equation}
Neither $\Pket$ nor $\Rket$ is an energy eigenstate for finite $N$,
and their overlap $S=\langle P|R\rangle=(1-m_*^2)^{N/2}$ is
exponentially small in the ordered phase (at $\Gamma/J=0.95$,
$N=370$: $S=(\Gamma/J)^N=(0.9025)^{185}\approx5.7\times10^{-9}$), so treating them as orthogonal is an
excellent approximation throughout this paper.
The true eigenstates are well approximated by the symmetric and antisymmetric superpositions
\begin{equation}
  |E_0\rangle = \frac{\Pket+\Rket}{\sqrt{2+2S}}\;(\text{ground state}),
  \qquad
  |E_1\rangle = \frac{\Pket-\Rket}{\sqrt{2-2S}}\;(\text{first excited state}),
  \label{eq:eigenstates}
\end{equation}
separated by the tunnel splitting $\DeltaE(N)$.
The degenerate thermodynamic limit ($\DeltaE\to0$ as $N\to\infty$)
where $\Pket$ and $\Rket$ become true ground states is approached
exponentially; for finite $N$ in the Goldilocks zone, $|E_0\rangle$
is the unique ground state.

\begin{remark}[Validity of $T=0$ eigenstates with a finite-$T$ bath]
\label{rem:T0}
The Hamiltonian eigenstates $|E_0\rangle$, $|E_1\rangle$ and the
tunnel splitting $\DeltaE$ are computed by diagonalising $\hat{H}$
without a thermal bath --- a $T=0$ calculation.
This is justified because the thermal population of $|E_2\rangle$ (the lowest level
above the ground doublet) is $e^{-\DeltaE_{20}/k_BT}\approx
e^{-11.4}\approx1\times10^{-5}$ (using $\DeltaE_{20}/\DeltaE_{10}
=1+\DeltaE_{21}/\DeltaE_{10}\approx11.4$ from exact diagonalisation at $N=370$, $\Gamma/J=0.95$),
so finite-temperature corrections to the eigenstates and to $m_*$ are
negligible at $T=10$~nK.
Temperature enters the paper in only one place: the Goldilocks
condition $\DeltaE=k_BT$ (which sets the operating point $N_c$).
The dephasing rate $\gamma_\phi$ is treated here as independently
controlled by the ambient magnetic field noise environment rather than
by the gas temperature; it does not modify the eigenstates or the
Hamiltonian spectrum used in the Lindblad simulation.
\end{remark}
The Bloch precession frequency is $\omega_0=2Jm_*$ and the
mean-field susceptibility $\chi_{\rm MF}=(\partial m/\partial h)_{h=0}
\sim(J-\Gamma)^{-1}\to\infty$ near criticality.

For finite $N$, the $\mathbb{Z}_2$ symmetry at $h=0$ is preserved:
the true ground state is a superposition $\alpha\Pket+\beta\Rket$
separated from the first excited state by the tunnel splitting
$\DeltaE(N)$.
Projecting the longitudinal bias into the $\{\Pket,\Rket\}$ pseudospin basis yields the effective two-level Hamiltonian $H_{\rm eff}=-Nm_*h\hat\tau_z-(\DeltaE/2)\hat\tau_x$. Differentiating the ground-state magnetisation $m_z$ with respect to $h$ at $h=0$ gives the intensive finite-$N$ susceptibility (which scales as $\sim N$ at the Goldilocks crossover where $\Delta E \sim k_B T$ is fixed):
 \begin{equation}
\chi_{\rm eff} = \frac{2Nm_*^2}{\DeltaE_{\rm tunnel}(N,\Gamma)},
  \label{eq:chi_eff}
\end{equation}
which is the two-level estimate. At $N=370$, $\Gamma/J=0.95$, exact
diagonalisation gives $\chi_{\rm eff}^{\rm exact}\approx1509~J^{-1}$
versus the two-level estimate $2Nm_*^2/\DeltaE\approx2049~J^{-1}$, a
35\% overestimate from multi-level contributions.

\section{The \texorpdfstring{$N=2$}{N=2} Exact Solution}
\label{sec:n2}

The $N=2$ case is the minimal exactly solvable instance: it proves
algebraically that $\Pket$ and $\Rket$ coexist coherently in the exact
quantum ground state for any $\Gamma>0$, demonstrating coherent superposition
of the two ordered phases and anchoring the large-$N$ instanton picture.
(Note: this establishes coherent superposition in the quantum ground state,
not a strict no-go theorem against all classical stochastic descriptions of
individual observables.)

At $N=2$, $h=0$, the symmetric Dicke subspace splits into even- and
odd-parity sectors.
The odd-parity sector contains the single eigenstate
$|a\rangle=(\Pket_2-\Rket_2)/\sqrt{2}$
(where $\Pket_2=|{+}{+}\rangle$, $\Rket_2=|{-}{-}\rangle$),%
\footnote{At $N=2$ we use the Dicke basis extremes
$\Pket_2=|{+}{+}\rangle$ and $\Rket_2=|{-}{-}\rangle$, which are
exactly orthogonal ($\langle P_2|R_2\rangle=0$).
For general $N$ and finite $\Gamma$, $|P\rangle$ and $|R\rangle$ are
spin-coherent states centred at $\pm m_*<1$, with exponentially small
overlap $S=(1-m_*^2)^{N/2}$ (at $N=370$, $\Gamma/J=0.95$:
$S\approx5.7\times10^{-9}$, so the orthogonality used here is an excellent
approximation).
Both descriptions identify the same physical objects --- the two
macroscopic ordered-phase minima --- with the Dicke-state labelling
being exact for $N=2$ and the spin-coherent-state picture being the
appropriate large-$N$ description.}
with energy $-J$, independent of $\Gamma$.
It is \emph{not} the overall ground state: for any $\Gamma>0$ the
even-parity level $E_-=-\frac{1}{2}(J+\sqrt{J^2+16\Gamma^2})<-J$
lies below $-J$, so $|GS\rangle$ (the ground state) is always even-parity.
The even-parity sector is spanned by
$|s\rangle=(\Pket_2+\Rket_2)/\sqrt{2}$ and
$|0\rangle=(|{+}{-}\rangle+|{-}{+}\rangle)/\sqrt{2}$, with eigenvalues
$E_\pm=\frac{1}{2}(-J\pm\sqrt{J^2+16\Gamma^2})$.
The ground state:
\begin{equation}
  |GS\rangle = \cos\tfrac{\phi}{2}\,|s\rangle
             + \sin\tfrac{\phi}{2}\,|0\rangle,
  \qquad \tan\phi = \frac{4\Gamma}{J}.
  \label{eq:gs2}
\end{equation}
The physical tunnel splitting is the gap between $|a\rangle$ and $E_-$:
\begin{equation}
  \DeltaE_{N=2} = \tfrac{1}{2}\!\left(-J+\sqrt{J^2+16\Gamma^2}\right)
    \approx \frac{4\Gamma^2}{J} \quad (\Gamma\ll J).
  \label{eq:split_n2}
\end{equation}
This algebraic (not exponentially suppressed) splitting confirms that
$\Pket$ and $\Rket$ are simultaneously present in the ground state for
any $\Gamma>0$, which is the basic mechanism behind mesoscopic quantum
coherence in the LMG model.
The combined $\Pket$/$\Rket$ weight in the even-parity ground state is
$P_{\rm macro}\equiv|\langle s|GS\rangle|^2
= \cos^2(\phi/2)$, which is always positive and approaches unity
deep in the ordered phase ($J\gg\Gamma$).

\section{Mean-Field Theory and Classical Dynamics}
\label{sec:mf}
In the $N\to\infty$ limit, the Bloch vector $\mathbf{m}=(m_x,m_y,m_z)$
obeys purely reactive mean-field (precessional) dynamics
(derived in Appendix~\ref{app:mf}):
\begin{equation}
  \dot{m}_z = -2\Gamma m_y,\quad
  \dot{m}_y = -2(Jm_z+h)m_x+2\Gamma m_z,\quad
  \dot{m}_x = 2(Jm_z+h)m_y.
  \label{eq:bloch}
\end{equation}
These equations capture the coherent classical limit: they are
conservative, support persistent oscillations about the ordered minima
$m_z=\pm m_*$, and admit no thermal activation across the free-energy
barrier.
To model barrier crossing as a classical foil, we coarse-grain the
fast transverse oscillations and treat the macroscopic longitudinal
dynamics via a thermodynamically consistent overdamped Langevin model
(Model~A dynamics, Appendix~\ref{app:fp})~\cite{Hohenberg1977}.
Model~A is not the decohered limit of the quantum BEC; it is a
separate phenomenological description of classical thermal activation
over the barrier $\Delta f_0$, introduced specifically as a foil
against which the quantum Landau--Zener crossover is compared.
The resulting effective Fokker--Planck equation is:
\begin{equation}
  \frac{\partial P}{\partial t} = \frac{\partial}{\partial m_z}
    \!\left[\Gamma_{\rm eff}\frac{\partial f(m_z)}{\partial m_z}P
    + D\frac{\partial P}{\partial m_z}\right],
  \label{eq:fp}
\end{equation}
where $\Gamma_{\rm eff}$ is an effective phenomenological mobility
and $D=\Gamma_{\rm eff}k_BT/N$ is the diffusion coefficient
(Einstein relation; the $1/N$ scaling ensures the stationary
distribution satisfies equilibrium statistical mechanics for the
intensive variable $m_z$).
The free-energy landscape $f(m_z)$ entering Eq.~\eqref{eq:fp} is the
mean-field variational free energy (Eq.~\eqref{eq:freeenergy}), which
is a static thermodynamic object derived from the partition function
$Z=\mathrm{Tr}[e^{-\beta\hat{H}}]$; its double-well shape is
independent of the dynamical regime and remains valid as the
thermodynamic landscape for classical activation regardless of the
degree of dephasing.
The explicit derivation of the drift and diffusion coefficients from
the thermodynamically consistent Langevin equation is given in
Appendix~\ref{app:fp}; the key physical conclusion is that the
classical Kramers escape time scales as
$\tau_K\propto\Gamma_{\rm eff}^{-1}\,e^{N\Delta f_0/k_BT}$,
exponentially large in $N$ regardless of whether the classical
dynamics is overdamped or underdamped (the prefactor $\Gamma_{\rm eff}^{-1}$
differs between regimes; the exponential does not~\cite{Kramers1940}).
The factor $N$ in the Boltzmann exponent is essential --- it is the
extensive free energy that governs the stationary distribution
$P_{\rm stat}(m_+)/P_{\rm stat}(m_-)=e^{2Nm_*h/k_BT}$
(where $m_\pm\equiv\pm m_*$ denote the two ordered minima),
encoding the $N$-amplified bias sensitivity between $\Pket$ and $\Rket$.
The mean first-passage time approaches the finite Kramers time
$\tau_K\propto\exp(N\Delta f_0/k_BT)$ as $h\to0$
(Appendix~\ref{app:mfpt}), where
$\Delta f_0\equiv f(0)-f(m_*)>0$ is the mean-field free-energy barrier
per spin between the saddle point $m_z=0$ and the ordered minimum
$m_z=m_*$, evaluated from Eq.~\eqref{eq:freeenergy} at $h=0$.
\section{The Goldilocks Zone}
\label{sec:goldilocks}

In the large-$N$ ordered phase, the tunnel splitting between $\Pket$ and
$\Rket$ is exponentially suppressed:
\begin{equation}
  \DeltaE(N) = C_0\,N^{1/2}\,e^{-NS_{\rm inst}},
  \qquad
  S_{\rm inst} = \operatorname{arctanh}(m_*)-m_*,
  \label{eq:tun}
\end{equation}
where the instanton prefactor $C_0$ is determined by fitting
Eq.~\eqref{eq:tun} to exact diagonalisation data in the range $N \in [50, 500]$
(see Appendix~\ref{app:instanton} for the physical origin of the $N^{1/2}$
zero-mode prefactor).
The exponent $S_{\rm inst}$ is derived semiclassically via the WKB instanton method;
the prefactor $C_0$ is empirically calibrated and is not predicted by the
semiclassical theory.
The fitted values $C_0/k_BT=2.51,\,2.546,\,2.616$ for
$\Gamma/J=0.99,\,0.95,\,0.90$ respectively are used throughout
(yielding $\ln(C_0/k_BT)=0.920,\,0.934,\,0.962$), and $C_0\sim\hbar\omega_0=2\hbar Jm_*$
sets the natural scale (derivation of $S_{\rm inst}$ in
Appendix~\ref{app:instanton}).
The Goldilocks crossover $N_c$ is defined by $\DeltaE(N_c)=k_BT$.
Dropping the $N^{1/2}$ prefactor gives the analytic estimate:
\begin{equation}
  N_c^{\rm analytic} \approx \frac{\ln(C_0/k_BT)}{S_{\rm inst}}.
  \label{eq:nc}
\end{equation}
This approximation is valid only when $N_cS_{\rm inst}\gg\tfrac{1}{2}\ln N_c$.
For all three parameter sets in Table~\ref{tab:goldilocks}, the left- and
right-hand sides are of comparable magnitude (ratio $\approx1.2$--$1.4$),
so the condition fails to hold: the analytic formula systematically
underestimates $N_c$ by factors of $3.4$--$5.7$.
For $\Gamma/J=0.99$ the failure is most severe: the analytic estimate
($N_c\approx972$) underestimates the instanton root ($N_c\approx5521$)
by a factor of $5.7$.
The analytic formula should never be used for $\Gamma/J\gtrsim0.98$;
exact diagonalisation is mandatory for experimental planning.

\begin{remark}
One condition is required for the Goldilocks zone to exist:
$\hbar\omega_0>k_BT$ (where $\omega_0=2Jm_*$;
$\omega_0>1310$~rad/s at $T=10$~nK).
This is simultaneously the condition for the tunnel splitting to be
thermally resolvable and for the analytic formula Eq.~\eqref{eq:nc}
to yield a positive $N_c$ estimate (since $\ln(C_0/k_BT)>0$ requires
$C_0\sim\hbar\omega_0>k_BT$).
In practice the analytic formula Eq.~\eqref{eq:nc} is only \emph{useful}
when $\ln(C_0/k_BT)\gg\tfrac{1}{2}\ln N_c$; below $\omega_0\sim100$~rad/s
this fails entirely ($\ln\approx-2.6<0$) and the zone does not exist.
For BEC experiments, Feshbach resonance tuning and tight trapping achieve
$\omega_0\sim5000$~rad/s comfortably within this window~\cite{Trenkwalder2016}.
\end{remark}

\begin{table}[H]
\centering
\small
\caption{Goldilocks crossover $N_c$ for BEC parameters ($T=10$~nK),
computed using the exact theoretical WKB action $S_{\rm inst}$ and the
fitted prefactors $C_0/k_BT=2.51,\,2.546,\,2.616$ for
$\Gamma/J=0.99,\,0.95,\,0.90$ respectively
($\ln(C_0/k_BT)=0.920,\,0.934,\,0.962$).
``Instanton root'' solves $C_0N_c^{1/2}e^{-N_cS_{\rm inst}}=k_BT$
numerically using the fitted $C_0$ values; it carries $\mathcal{O}(1)$
uncertainty from the instanton approximation and should be validated by
exact diagonalisation.
Uncertainty estimates $\pm\delta N_c$ are obtained by varying $C_0$ by
$\pm5\%$ (the typical fitting uncertainty) and solving for the new root.
For $\Gamma/J=0.95$, exact diagonalisation gives $\DeltaE/k_BT=1.000$
at $N=370$, confirming the instanton root $360\pm50$ is consistent.
$^\dagger$The analytic approximation (which drops the $N^{1/2}$ prefactor)
uses the same fitted $C_0$ values and is shown only to quantify the
approximation error; it should \emph{never} be used for experimental
planning.
The validity condition $N_cS_{\rm inst}\gg\tfrac{1}{2}\ln N_c$ fails for
all three cases (left- and right-hand sides are comparable, ratio
$\approx1.2$--$1.4$), which is why the analytic formula underestimates
$N_c$ by factors of $3.4$--$5.7$.}
\label{tab:goldilocks}
\setlength{\tabcolsep}{8pt}
\begin{tabular}{@{}ccccc@{}}
\toprule
$\Gamma/J$ & $S_{\rm inst}$ & $\ln(C_0/k_BT)$ &
$N_c$ Analytic$^\dagger$ & $N_c$ Instanton root\\
\midrule
$0.99$ & $0.000947$ & $0.920$ &
  $\sim972\;\;(\times5.7$ error$)$ & $\sim5521\pm\mathcal{O}(1000)$\\
$0.95$ & $0.010787$ & $0.934$ &
  $\sim87\;\;\;\;(\times4.1$ error$)$ & $\sim360\pm50$ \\
$0.90$ & $0.031255$ & $0.962$ &
  $\sim31\;\;\;\;(\times3.4$ error$)$ & $\sim105\pm15$ \\
\bottomrule
\end{tabular}
\end{table}
\section{Open Quantum Dynamics}
\label{sec:lindblad}

Coupling to a Markovian environment gives the
Gorini--Kossakowski--Sudarshan--Lindblad (GKSL) master equation:
\begin{equation}
  \dot\rho = -i[\hat{H},\rho]
    + \sum_k\!\left(L_k\rho L_k^\dagger
    - \tfrac{1}{2}L_k^\dagger L_k\rho
    - \tfrac{1}{2}\rho L_k^\dagger L_k\right).
  \label{eq:lindblad}
\end{equation}

\begin{table}[H]
\centering
\small
\caption{Summary of physical timescales and energy scales.}
\label{tab:timescales}
\begin{tabular}{@{}ll@{}}
\toprule
Symbol & Definition \\
\midrule
$\DeltaE$ & Quantum tunnel splitting (energy gap between $|E_0\rangle$
            and $|E_1\rangle$). \\
$T_1$     & Longitudinal population relaxation time (spin thermalisation). \\
$T_2^{\rm cl}$   & Classical transverse relaxation time ($\approx 1/\gamma_\phi$). \\
$T_2^{\rm coll}$ & Mean-field collective coherence time
                   (phenomenological two-level estimate). \\
$T_2^{\rm phys}$ & Exact eigenstate Liouvillian decay time of the
                   $|E_0\rangle\langle E_1|$ coherence. \\
$\tau_K$  & Classical Kramers escape time (macroscopic barrier crossing). \\
$\tau_Q$  & Quench time (duration of the experimental bias-field sweep). \\
\bottomrule
\end{tabular}
\end{table}

\begin{remark}[$T_2$ notation]
Three distinct coherence times appear in this paper.
$T_2^{\rm cl}\equiv1/\Gamma_2$ is the classical transverse relaxation time
appearing in heavily damped Bloch equations, set by the ambient magnetic
field noise via
$1/T_2^{\rm cl}=\gamma_\phi+1/(2T_1)\approx\gamma_\phi$
(since $T_1\gg T_2^{\rm cl}$ in spinor BECs~\cite{Trenkwalder2016}).
$T_2^{\rm coll}$ (Eq.~\eqref{eq:n2_rate}) is the mean-field estimate of
the $\Pket$--$\Rket$ quantum coherence time, derived by treating
$\Pket$, $\Rket$ as spin-coherent states.
$T_2^{\rm phys}$ (Eq.~\eqref{eq:T2phys}) is the exact Liouvillian decay
time of the $|E_0\rangle\langle E_1|$ energy-eigenstate coherence,
computed from exact matrix elements; this is the authoritative
experimental threshold.
\end{remark}

\begin{remark}[Physical origin of collective dephasing in spinor BECs]
A uniform stochastic magnetic field $\delta B(t)$ along the quantisation
axis couples to the collective operator $\hat{J}_z$ via the Zeeman term,
yielding the collective jump operator $L_z=\sqrt{\gamma_\phi}\hat{J}_z$.
This noise dominates over local inhomogeneous broadening in
well-shielded BEC experiments.
The physical dephasing rate $\gamma_\phi$ corresponds to the pure
dephasing contribution to the classical transverse relaxation rate:
$1/T_2^{\rm cl} = \gamma_\phi + 1/(2T_1) \approx \gamma_\phi$.

The quantum Lindblad dephasing and the
classical thermal activation (Section~\ref{sec:mf}) are driven by
physically distinct environments.
Classical Kramers activation is governed by the $T=10$~nK motional bath
of the BEC gas, which obeys the fluctuation-dissipation theorem at that
temperature --- this is precisely what gives the Einstein relation
$D=\Gamma_{\rm eff}k_BT/N$ its physical meaning and ensures the
stationary distribution is the correct Boltzmann factor
$e^{-N\Delta f_0/k_BT}$.
The quantum dephasing bounded here is instead driven by technical
magnetic-field noise acting as a \emph{pure-dephasing} environment
with $T_1\to\infty$ --- an infinite-temperature bath that introduces
phase fluctuations with no energy relaxation and does not restore
detailed balance.
Conflating these two environments would render the Einstein relation
physically meaningless and collapse the Kramers barrier.

The Markovian (white-noise) Lindblad model for the magnetic-field noise
is an approximation: real BEC magnetic field noise has $1/f$ spectral
components and is non-Markovian.
Non-Markovian dephasing (with memory) can decohere more slowly than
Markovian white noise in many experimentally relevant low-frequency-noise
models (such as $1/f$ magnetic-field noise in BECs), though this is not
a general theorem for all non-Markovian environments.
Under such low-frequency-dominated noise, the Markovian threshold
$\gamma_\phi\lesssim0.289$~s$^{-1}$ derived below provides a
\emph{reference benchmark} for the permissible dephasing rate;
whether the true non-Markovian threshold is more or less permissive
depends on the precise spectral density and protocol filter function,
though for $1/f$-dominated noise the Markovian bound is often
conservative~\cite{Emary2014,Paladino2014}.
\end{remark}

\begin{remark}[Local vs.\ collective dephasing: $N$-scaling]
\label{rem:dephasing}
Full derivations are in Appendix~\ref{app:decoherence}.
For local baths ($L_z^{(i)}=\sqrt{\gamma_\phi}\hat\sigma_i^z$),
the local-bath coherence time $T_2^{\rm local}$ satisfies:
$1/T_2^{\rm local}=N\gamma_\phi(1+m_*^2)$.
For a collective bath ($L_z=\sqrt{\gamma_\phi}\hat{J}_z$), the
collective-bath coherence time $T_2^{\rm coll}$ governing the
$\Pket$--$\Rket$ coherence decays as:
\begin{equation}
  \frac{1}{T_2^{\rm coll}} \approx \frac{m_*^2N^2\gamma_\phi}{2}.
  \label{eq:n2_rate}
\end{equation}
\emph{Note:} $T_2^{\rm coll}$ is a mean-field estimate using the
coherent-state approximation for $\Pket,\Rket$; the exact eigenstate
coherence time $T_2^{\rm phys}$ (Eq.~\eqref{eq:T2phys}, derived from
full $371$-dimensional diagonalisation) is the more reliable quantity
and enters the Lindblad simulation directly.
At $\gamma_\phi=0.05$~s$^{-1}$: $T_2^{\rm coll}\approx3.0$~ms versus
$T_2^{\rm phys}\approx7.04$~ms --- a factor of $2.35$ arising from
the parity cross-term elimination detailed in Remark~\ref{rem:reconcile}.
The collective $N^2$ scaling of $T_2^{\rm coll}$ explains why the
Goldilocks zone is intrinsically narrow: as $N$ grows beyond $N_c$,
$T_2^{\rm coll}$ drops quadratically, collapsing the quantum coherence
between $\Pket$ and $\Rket$ before macroscopic susceptibility is useful.
At $N=370$, $\gamma_\phi=0.05$~s$^{-1}$:
$T_2^{\rm coll}\approx3.0$~ms (collective) versus
$T_2^{\rm local}\approx49$~ms (local).
The collective bath is the relevant model for BEC global field noise.

The chain connecting microscopic dephasing to the LGI threshold passes
through three levels (see Remark~\ref{rem:hierarchy}):
$\gamma_\phi \xrightarrow{\text{level A}}
T_2^{\rm coll}$ (Eq.~\eqref{eq:n2_rate},
$\gamma_\phi^{\rm (A)}\lesssim0.050$~s$^{-1}$,
mean-field two-level formula, most conservative)
$\xrightarrow{\text{level B}} T_2^{\rm phys}$ (Eq.~\eqref{eq:T2phys},
$\gamma_\phi^{\rm (B)}\lesssim0.117$~s$^{-1}$, exact eigenstate two-level,
$2.35\times$ improvement from parity cross-term elimination)
$\xrightarrow{\text{level C}}$ five-level Lindblad simulation
($\gamma_\phi^{\rm (C)}\lesssim0.289$~s$^{-1}$, authoritative experimental
threshold, $2.47\times$ further improvement from multi-level Dicke-doublet
structure).
\end{remark}

\begin{remark}[$N=370$ as a benchmark]
The value $N=370$ is chosen as the exact-diagonalisation crossover
where the numerically exact tunnel splitting equals $k_BT$
(verified: $\DeltaE=1310$~rad/s at $N=370$, $\Gamma/J=0.95$).
This is distinct from the instanton-root crossover $N_c^{\rm inst}\approx360$
(Table~\ref{tab:goldilocks}), which solves the instanton formula
$C_0N^{1/2}e^{-NS_{\rm inst}}=k_BT$ using the fitted prefactor
$C_0/k_BT=2.546$ and carries $\mathcal{O}(1)$ prefactor uncertainty.
The value $N=370$ is a natural benchmark but not the experimentally
optimal size; the Goldilocks double-squeeze
(Section~\ref{sec:disc}) restricts the optimal range to
$N\approx250$--$300$ where both the LGI violation margin and the
Goldilocks coherence condition are simultaneously satisfied with margin.
A full parameter sweep using exact diagonalisation
(Appendix~\ref{app:code}) confirms this optimal balance.
\end{remark}
\section{Falsifiable Predictions}
\label{sec:predictions}

\begin{table}[H]
\centering
\small
\caption{Falsification matrix. P1--P3 are not quantum-distinctive
(reproducible by classical bistable systems). P4 and P5 are the
genuine discriminators.
Abbreviations: ACF = autocorrelation function;
KZM = Kibble--Zurek Mechanism (universal defect-formation scaling
under a critical quench~\cite{Defenu2018,Puebla2020,Kopylov2019}).
For P3: the J-quench exponent $\tau_Q^{1/3}$ is standard KZM
(genuine quantum critical point, diverging correlation length;
$z=1$, $\nu=1/2$, freeze-out $\hat{t}\propto\tau_Q^{z\nu/(z\nu+1)}=\tau_Q^{1/3}$).
The h-quench exponent $\tau_Q^{1/2}$ is LZ bias-sweep scaling
(avoided level crossing in the already-ordered phase, not a critical
phenomenon). In classical dynamics, the $\tau_Q^{1/2}$ scaling arises
from overdamped continuous (pitchfork) bifurcation scaling with
dynamical exponent $\mu=1$ ($\tau_{\rm rel}\propto|\varepsilon|^{-1}$,
giving $\hat{t}\propto\tau_Q^{1/2}$) --- distinct from the spinodal
(saddle-node) scaling $\mu=1/2$ which gives $\hat{t}\propto\tau_Q^{1/3}$.
Both the quantum LZ interaction time and the classical pitchfork
freeze-out scale as $\tau_Q^{1/2}$ for distinct physical reasons.
For the quantum h-quench, the $\tau_Q^{1/2}$ timescale governs the
LZ Zener interaction time $t_Z\sim\sqrt{\hbar\tau_Q/(Nm_*\Delta h)}$;
however, the quantum \emph{outcome} is an exponential
$P_{\rm error}\propto e^{-{\rm const}\cdot\tau_Q}$, not a power-law
defect density. The $\tau_Q^{1/2}$ entry in this table therefore refers
to the classical scaling analogy only.}
\label{tab:falsification}
\setlength{\tabcolsep}{6pt}
\begin{tabular}{@{}lccc@{}}
\toprule
Prediction & Class.\ bistable & Quantum LMG & Q-distinctive?\\
\midrule
P1: critical slowing  & div.\ at $\Gamma{=}J$ (mean-field)  & diverges        & No\\
P2: oscillatory ACF   & yes                     & yes             & No\\
P3a: KZM ($J$-quench) & $\tau_Q^{1/3}$ (underdamped) & $\tau_Q^{1/3}$ & No\\
P3b: LZ bias-sweep ($h$-quench) & $\tau_Q^{1/2}$ (pitchfork, overdamped)
                                & $\tau_Q^{1/2}$ & No\\
P4: LZ error rate     & $\to1$ (stuck)          & $\to0$          & \textbf{Yes}\\
P5: LGI $K_3$         & $\leq1$                 & $\leq1.415$     & \textbf{Yes}\\
\bottomrule
\end{tabular}
\end{table}

\subsection{Classical diagnostics (P1--P3)}

\textbf{P1.} The mean first-passage time diverges only at $\Gamma=J$;
in the ordered phase it approaches the Kramers time
$\tau_K\propto\Gamma_{\rm eff}^{-1}\,e^{N\Delta f_0/k_BT}$
(where $\Delta f_0\equiv f(0)-f(m_*)>0$ is the mean-field free-energy
barrier per spin; see Appendix~\ref{app:mfpt}).
At our benchmark parameters ($N=370$, $\Gamma/J=0.95$, $T=10$~nK),
the barrier evaluates to $N\Delta f_0/k_BT\approx13.1$
(derivation in Appendix~\ref{app:mfpt}),
giving $\tau_K\sim\mathcal{O}(10^2)$~s.
Crucially, the $N$-dependence is confined strictly to the exponential:
the potential curvature factor $\propto1/N$ in the Kramers prefactor is
exactly cancelled by $D\propto1/N$ from the Einstein relation,
leaving $N$ exclusively in the Boltzmann exponent.

\textbf{P2.} In the underdamped regime, the normalised autocorrelation
function of $m_z$ is
$C(\Delta t)=A_0\,e^{-\Gamma_2\Delta t}\cos(\omega_0\Delta t)
  + B_0\,e^{-\Gamma_1\Delta t}$,
where $A_0$ is the oscillatory amplitude, $B_0$ is the slow longitudinal
baseline (negligible for $\Delta t\ll T_1$),
$\omega_0=2Jm_*$ is the normal-mode oscillation frequency about the
ordered minimum (derived by linearising the Bloch equations around
the tilted fixed point, Appendix~\ref{app:acf}),
and $\Gamma_2\equiv1/T_2^{\rm cl}$ is the transverse decoherence rate
($\Gamma_1\equiv1/T_1$).
This provides direct experimental access to $Jm_*$ (via $\omega_0=2Jm_*$)
and, given independent knowledge of $T_1$ and $T_2$, to $m_*$.

\textbf{P3.} A linear quench of $J$ gives freeze-out time
$\hat{t}\propto\tau_Q^{1/3}$, where $\hat{t}$ is the Kibble--Zurek
impulse time at which the system falls out of adiabaticity
(LMG universality class, dynamical exponent $z=1$, correlation-length
exponent $\nu=1/2$, giving $\hat{t}\propto\tau_Q^{z\nu/(z\nu+1)}
=\tau_Q^{1/3}$)~\cite{Defenu2018,Puebla2020,Kopylov2019}.
Classical bistable systems driven through a bifurcation also exhibit
freeze-out scaling: overdamped dynamics give $\tau_Q^{1/2}$
and underdamped dynamics give $\tau_Q^{1/3}$.
Because classical underdamped dynamics share the $\tau_Q^{1/3}$ exponent
with the quantum LMG, the J-quench alone is \emph{not} uniquely
quantum-distinctive.

A quench of $h$ gives $\hat{t}\propto\tau_Q^{1/2}$: for
$h\gg\DeltaE/(2Nm_*)\approx5.7$~rad/s (satisfied throughout the
experimental sweep window), the quantum gap grows linearly,
$\Delta(h)\approx2Nm_*h$, with dynamical exponent $\mu_h=1$, yielding
the $\tau_Q^{1/2}$ Zener interaction time
$t_Z\sim\sqrt{\hbar\tau_Q/(Nm_*\Delta h)}$.
In classical dynamics, the spinodal (saddle-node) bifurcation at
$h=h_{\rm sp}$ has $\mu=1/2$ ($\tau_{\rm rel}\propto|\varepsilon|^{-1/2}$)
and gives $\hat{t}\propto\tau_Q^{1/3}$.
However, the experimental sweep window (Eq.~\eqref{eq:sweep_window})
explicitly requires $\Delta h\ll h_{\rm sp}$: the bias sweep is
bounded away from the spinodal, so the classical system never approaches
the saddle-node bifurcation during the protocol.
Within this window, the classical response is governed by small
deviations from the occupied minimum, where the restoring force is
linear in the displacement ($\mu=1$), producing $\tau_Q^{1/2}$ scaling
--- the same as the overdamped pitchfork foil.
The spinodal is therefore not the correct classical comparison for the
P3b protocol as specified; the pitchfork ($\mu=1$) is the appropriate
foil, and the $\tau_Q^{1/2}$ exponent is shared by both quantum and
classical dynamics, making P3b non-discriminating on the basis of
freeze-out exponent alone.
Either way, the quantum \emph{outcome} of the h-quench ---
$P_{\rm error}\propto e^{-{\rm const}\cdot\tau_Q}$ from the LZ formula
--- is exponential in $\tau_Q$, not a power law, and has no classical
analogue in the kinetically frozen regime.
Kibble--Zurek power-law scaling alone therefore cannot discriminate
classical from quantum dynamics here, which is precisely why P4 and P5
are necessary.
\subsection{P4: Landau--Zener adiabatic error rate (Proposed Discriminator)}

\textbf{Scope of the classical comparison.}
The classical prediction $P_{\rm error}^{\rm class} \to 1$ is derived from a
phenomenological, non-ergodic activated model (Model~A / Fokker--Planck
dynamics, Section~\ref{sec:mf} and Appendix~\ref{app:fp}) rather than a
rigorous microscopic classical limit of the LMG Hamiltonian.
It serves as a well-defined coarse-grained foil representing macroscopic
thermal barrier crossing.
While this establishes a parametrically large quantum--classical separation
under this specific but broad class of classical kinetics, P4 is presented
here as a proposed discriminator, complementing the strictly bounded LGI
violation of P5.

\textbf{Classical non-ergodic limit.}
In spinor BEC experiments, spin thermalisation operates on timescales
$T_1\sim\text{seconds}$~\cite{Trenkwalder2016}, far exceeding any
experimentally accessible quench time $\tau_Q\sim\text{ms}$
(a free experimental parameter set by the laser detuning sweep rate).
The classical Kramers escape time
$\tau_K\propto\Gamma_{\rm eff}^{-1}\,e^{N\Delta f_0/k_BT}$
is exponentially large: at our benchmark parameters
$N\Delta f_0/k_BT\approx13$, so $\tau_K\sim\mathcal{O}(10^2)$~s
(see the spinodal remark below for the explicit estimate).
On any experimentally accessible quench time $\tau_Q\sim\text{ms}$,
the classical system remains effectively non-ergodic.
While the classical system rapidly equilibrates locally within its initial
well ($\tau_{\rm intrawell}\ll\tau_Q$), it cannot cross the macroscopic
barrier because $\tau_K\gg\tau_Q$.
It therefore remains frozen in the diabatic minimum throughout the sweep,
yielding
\begin{equation}
  P_{\rm error}^{\rm class}(\tau_Q\ll\tau_K) \to 1,
  \label{eq:floor}
\end{equation}
for all experimentally accessible sweep times.
This is the opposite of the quantum result: in the coherent regime
$\hbar/\DeltaE\ll\tau_Q\ll T_2^{\rm phys}$ (sweep long enough for
adiabatic tunnelling but short enough that decoherence is negligible),
the quantum LMG system tunnels coherently from $\Pket$ to $\Rket$ with
exponentially decreasing error (Eq.~\eqref{eq:lz}).
The quantum--classical separation is therefore \emph{parametrically
large}: classical $P_{\rm error}\to1$ versus quantum
$P_{\rm error}\to0$, making this a robust discriminator that does not
rely on precise knowledge of temperature.

\begin{remark}[Why the Boltzmann-floor argument does not apply here]
One might expect the classical floor to be the thermal equilibrium
population $(1+e^{\DeltaE/k_BT})^{-1}$ of the higher-energy well.
This applies only to ergodic classical systems that thermalise on a
timescale far shorter than $\tau_Q$.
The BEC is explicitly non-ergodic on the quench timescale
($\tau_K\gg\tau_Q$ due to the exponential barrier factor),
so the equilibrium Boltzmann distribution is never reached classically.
Furthermore, $\DeltaE$ is the quantum tunnel splitting --- there is no
corresponding classical energy splitting between the two wells at $h=0$,
since the classical system has two exactly degenerate minima there.
Importing $\DeltaE$ into a classical floor formula would conflate quantum
and classical physics.
\end{remark}

\begin{remark}[Two-level approximation near criticality]
\label{rem:two_level}
At $\Gamma/J=0.95$, $N=370$: exact diagonalisation gives
$\DeltaE_{10}=0.0352\,J$, $\DeltaE_{21}=0.365\,J$, so
$\DeltaE_{21}/\DeltaE_{10}\approx10.4$.
The first gap is anomalously small near the instanton crossover; the
five-level truncation is converged to $\pm5\%$ (Remark~\ref{rem:conv}).
The LZ and LGI formulae are therefore \emph{qualitatively} predictive
near criticality, with order-unity corrections from multi-level effects
quantified in Section~\ref{sec:multilevel}.
\end{remark}

\textbf{Quantum coherent LZ formula.}
For an $h$-quench through the $\Pket$--$\Rket$ degeneracy, the effective
two-level Hamiltonian in the $\{\Pket,\Rket\}$ pseudospin basis is
\begin{equation}
  H = -Nm_*v_h t\,\hat\tau_z - (\DeltaE/2)\hat\tau_x,
\end{equation}
where $\hat\tau_{x,z}$ are Pauli matrices in the $\{\Pket,\Rket\}$
two-level subspace, $v_h=\Delta h/\tau_Q$ is the sweep rate,
and $\Delta h$ is the total change in longitudinal bias $h$
(in rad\,s$^{-1}$ in natural units $\hbar=1$; in BEC notation
$\Delta h=\hbar\Delta\delta$ where $\Delta\delta$ is the laser detuning
change in rad\,s$^{-1}$).
Mapping to Zener's canonical form $H=At\,\hat\tau_z+V\hat\tau_x$ gives
$A=Nm_*v_h=\alpha/\tau_Q$ and $V=\DeltaE/2$
(the sign of the diagonal term is absorbed into the initial condition;
only $|A|$ enters the transition probability),
where $\alpha\equiv Nm_*\Delta h$ is independent of $\tau_Q$
(units: rad\,s$^{-1}$).
We assume the sweep range $\Delta h$ satisfies
Eq.~\eqref{eq:sweep_window}: large enough for the LZ interaction to
occur ($\Delta h\gg\DeltaE/(Nm_*)$) but small enough that the system
does not reach the spinodal ($\Delta h\ll h_{\rm sp}$), ensuring the
classical system remains trapped while the quantum system tunnels.

\begin{remark}[Spinodal constraint and valid sweep window]
\label{rem:spinodal}
The spinodal field $h_{\rm sp}$ at which the metastable well disappears
satisfies $f''(m_{\rm sp})=0$ simultaneously with $f'(m_{\rm sp})=0$.
Setting $y\equiv Jm+h$ and $E\equiv\sqrt{y^2+\Gamma^2}$, these conditions
give $E_{\rm sp}=(J\Gamma^2)^{1/3}$,
$y_{\rm sp}=\sqrt{(J\Gamma^2)^{2/3}-\Gamma^2}$,
$m_{\rm sp}=-y_{\rm sp}/E_{\rm sp}$,
and from $f'=0$:
\begin{equation*}
  h_{\rm sp} = y_{\rm sp}\!\left(\frac{J^{2/3}}{\Gamma^{2/3}}-1\right).
\end{equation*}
At $\Gamma/J=0.95$: $y_{\rm sp}\approx0.177\,J$,
$m_{\rm sp}\approx-0.183$, and
$h_{\rm sp}\approx0.0062\,J\approx229$~rad/s.
The minimum sweep amplitude for LZ physics is
$\DeltaE/(Nm_*)\approx11$~rad/s.
The valid window for P4 is therefore
\begin{equation}
  \frac{\DeltaE}{Nm_*}\approx11~\text{rad/s}
  \;\ll\; \Delta h \;\ll\;
  h_{\rm sp}\approx229~\text{rad/s},
  \label{eq:sweep_window}
\end{equation}
a factor-of-$\sim20$ controllable range, experimentally accessible via
the laser detuning sweep rate.
For $\tau_Q\sim\text{ms}$ and
$\tau_K\approx(2\pi/\omega_0)\,e^{N\Delta f_0/k_BT}
\approx0.27~\text{ms}\times e^{13.1}\approx130$~s
$\gg\tau_Q$ (barrier height derived in Section~\ref{sec:predictions}),
the classical system remains non-ergodic throughout.
Sweeps with $\Delta h>h_{\rm sp}$ carry the classical system past the
spinodal, causing the metastable well to vanish and giving
$P_{\rm error}\to0$ classically; the protocol must respect
Eq.~\eqref{eq:sweep_window} to maintain the quantum--classical
discrimination.
\end{remark}

The standard Zener formula~\cite{Zener1932,Landau1932}
$P=\exp(-\pi V^2/\hbar|A|)$ then yields:
\begin{equation}
  P_{\rm error}^{\rm quant}(\tau_Q)
    = \exp\!\left(-\frac{\pi\DeltaE^2\,\tau_Q}{4\hbar\alpha}\right)
    \xrightarrow{\tau_Q\to\infty}0.
  \label{eq:lz}
\end{equation}
The factor of 4 arises from $V=\DeltaE/2$ (so $V^2=\DeltaE^2/4$).
Unlike P5 (Section~\ref{sec:multilevel}), which is treated by full
numerical Lindblad simulation, P4 rests on the analytic LZ formula for
the coherent regime ($\tau_Q\ll T_2^{\rm phys}$) and the time-hierarchy
argument $\tau_K\gg\tau_Q$ for the classical non-ergodic regime; a
complete open-system treatment of the crossover region is left for
future work.
The coherent LZ formula~\eqref{eq:lz} is valid when decoherence is
negligible during the sweep, i.e.\ $\tau_Q\ll T_2^{\rm phys}$;
the lower bound $\tau_Q\gg\hbar/\DeltaE\approx0.76$~ms ensures the
sweep is slow enough for adiabatic tunnelling to occur.
For $\tau_Q\gtrsim T_2^{\rm phys}$ the coherent approximation fails
and the transition probability requires a first-principles open-system
LZ calculation~\cite{Fraas2017}.
The classical non-ergodic result $P_{\rm error}\to1$ applies for all
$\tau_Q\ll\tau_K$; since $\tau_K\gg T_2^{\rm phys}$ in spinor BECs,
there is a broad intermediate window
($T_2^{\rm phys}\lesssim\tau_Q\ll\tau_K$) where the system is neither
fully coherent nor classically ergodic --- the transition probability
in this window is given by the open-system quantum result, not by the
classical frozen limit.

\subsection{P5: Leggett--Garg Inequality violation}
\label{sec:p5}

\textbf{Observable.}
We use $Q=\operatorname{sgn}(\hat{J}_z)$, which takes value $+1$ when the
system is in the $\Pket$ phase and $-1$ when in the $\Rket$ phase.
For even $N$, the $m=0$ Dicke state is present in the basis and
$Q\vert_{m=0}=0$, making $Q$ technically trichotomic rather than dichotomic.
Strictly, the Leggett--Garg bound $K_3\leq1$ is derived for dichotomic observables.
One could adopt the convention $\operatorname{sgn}(0)=+1$ to restore dichotomic
status, as in Sakamoto et al.~\cite{Sakamoto2024}; however, this would introduce
a nonzero diagonal element $Q_{00}\approx+0.0024\neq0$, breaking the exact
$T_1$ immunity that is central to this paper.
We instead retain $Q\vert_{m=0}=0$ to preserve the exact parity structure.
The corrected macrorealist bound for a trichotomic observable whose middle value
$Q=0$ has weight $p_0=|\langle m{=}0|E_0\rangle|^2=0.24\%$ is
$K_3\lesssim1+2p_0\approx1.005$~\cite{Emary2014},
far below our violation of $K_3\approx1.317$, so the violation is unambiguous.
The entire $0.24\%$ weight (exact value: $|\langle m{=}0|E_0\rangle|^2=0.2374\%$,
from complete $371$-dimensional symmetric Dicke sector diagonalisation
at $N=370$, $\Gamma/J=0.95$) is already absorbed into the reduction
$Q_{01}^2=0.9436<1$ and propagated exactly through all calculations,
so it does not affect the LGI bound.

In the exact multi-level calculation, matrix elements
$Q_{ij}=\langle E_i|\operatorname{sgn}(\hat{J}_z)|E_j\rangle$ are
computed numerically in the Dicke basis. The LGI correlator is:
\begin{equation}
  C_{ij} = \tfrac{1}{2}\operatorname{Tr}\!\bigl[Q\,
    \mathcal{E}_{t_i\to t_j}(\{Q,\rho(t_i)\})\bigr],
  \qquad K_3 = C_{12}+C_{23}-C_{13},
  \label{eq:k3}
\end{equation}
where $\mathcal{E}_{t_i\to t_j}$ is the Lindblad channel,
$\rho(t_i)$ is the system state immediately before the measurement
at $t_i$, and $\{Q,\rho(t_i)\}$ is the anticommutator implementing the
state update after a non-selective projective measurement.
For the stationary protocol (re-preparation before each pair),
$\rho(t_1)=\rho(t_2)=\rho_0=|E_0\rangle\langle E_0|$;
for the true sequential protocol, $\rho(t_2)$ is the
post-measurement state after the $t_1$ measurement (see
Appendix~\ref{app:code} for the explicit simulation).

\begin{remark}[Robustness to thermal mixture in the ground doublet]
\label{rem:thermal}
At the Goldilocks point $\DeltaE=k_BT$, the thermal equilibrium
population of $|E_1\rangle$ is $1/(1+e)\approx26.9\%$.
One might therefore question whether $\rho_0=|E_0\rangle\langle E_0|$
is physically realistic.
Two points resolve this.
First, the spin state is \emph{prepared} before each LGI trial by
standard BEC ground-state protocols; since $T_1\sim\text{s}\gg\tau_{\rm exp}$,
the spin degree of freedom does not thermalise to the motional temperature
on the measurement timescale.
Second --- and more importantly --- the LGI correlator is \emph{exactly
immune} to the thermal mixture within the ground doublet.
For the thermal state
$\rho_{\rm th}=p_0|E_0\rangle\langle E_0|+p_1|E_1\rangle\langle E_1|$
(with $p_0+p_1\equiv1$ since higher levels have population
$\sim e^{-\DeltaE_{21}/k_BT}\approx3\times10^{-5}$, see Remark~\ref{rem:T0}),
the off-diagonal element of the instrument is:
\begin{equation}
  \tfrac{1}{2}\{Q,\rho_{\rm th}\}_{01}
    = Q_{01}\,\frac{p_0+p_1}{2} = \frac{Q_{01}}{2},
\end{equation}
identical to $\tfrac{1}{2}\{Q,\rho_0\}_{01}=Q_{01}/2$,
because $Q_{00}=Q_{11}=0$ exactly (collective $\mathbb{Z}_2$ parity).
Therefore $C_{\rm thermal}(t)=C_{\rm pure}(t)$ exactly within the doublet,
and $K_3$ is independent of the mixing ratio $p_0/p_1$.
This is not an approximation but an exact consequence of the parity
structure; the protocol does not require ground-state preparation
(higher levels are populated at $\lesssim10^{-5}$, see Remark~\ref{rem:T0}),
making it substantially more robust experimentally than a protocol
that relies on a specific population ratio.

This immunity is \emph{static}: it means the value of $K_3$ is
insensitive to the population ratio $p_0/p_1$ \emph{at the moment of
each measurement}.
It is distinct from dynamic $T_1$ decoherence: any $\hat{J}_\pm$-type
jump operator acting \emph{during} the free-evolution interval $\Delta t$
contributes $1/(2T_1)$ to the coherence decay rate via the standard
relation $1/T_2=1/(2T_1)+1/T_\phi$.
The protocol is therefore not immune to dynamic $T_1$ processes, but
these are negligible here because $T_1\sim\text{s}\gg\Delta t\approx0.8$~ms:
the correction to $1/T_2^{\rm phys}$ is $1/(2T_1)\lesssim0.5$~s$^{-1}$,
less than $0.4\%$ of the dominant pure-dephasing term $\gamma_\phi\Gamma_{01}
\approx142$~s$^{-1}$ at the BEC target.
\end{remark}
For a dichotomic observable with projectors $P_\pm=(I\pm Q)/2$, the
standard two-time correlator definition
$C_{12}=\sum_{q_1,q_2}q_1q_2\,P(q_1,q_2)$ gives directly
$\sum_{q_1}q_1 P_{q_1}\rho_0 P_{q_1}
 = P_+\rho_0 P_+ - P_-\rho_0 P_- = \frac{1}{2}\{Q,\rho_0\}$,
confirming that the anticommutator is the standard Lüders instrument for
any ideal projective LGI measurement (Emary et al.~\cite{Emary2014}, Eq.\ (8));
it is distinct from the non-selective post-measurement state
$\frac{1}{2}(\rho_0+Q\rho_0 Q)=P_+\rho_0 P_++P_-\rho_0 P_-$,
which does not appear in the correlator.
The non-selective instrument $\frac{1}{2}\{Q,\rho_0\}$ implements a
Lüders measurement~\cite{Emary2014} that averages over all outcomes
without selecting a definite result.
The post-measurement state is the proper mixture
$\rho_{\rm post}=P_+\rho_0 P_++P_-\rho_0 P_-$ --- an ensemble in
which each individual run has collapsed to a definite eigenstate of $Q$,
but the record of which outcome occurred is discarded.
The operator $\frac{1}{2}\{Q,\rho_0\}=P_+\rho_0 P_+-P_-\rho_0 P_-$
is not a density matrix (it is not positive semidefinite);
it is the operator-valued instrument needed to compute $C_{12}$,
and feeding it into the channel $\mathcal{E}$ before tracing with $Q$
implements the two-time correlator without assuming any particular
outcome at $t_1$.
No assumption of wavefunction collapse to a particular eigenstate
is built into the correlator formula, keeping the test free of
circularity with respect to the macrorealism hypothesis being tested.
Furthermore, because the ground-state wavefunction has negligibly small amplitude at the symmetric barrier $m=0$ (the $m=0$ Dicke-state weight is only $0.24\%$, see above), the projective measurement $Q = \operatorname{sgn}(\hat{J}_z)$ does not induce significant high-energy scattering (spectral leakage from the non-smooth sign operator). The post-measurement state carries $97.3\%$ of its weight in the ground doublet $\tfrac{1}{\sqrt{2}}(|E_0\rangle \pm |E_1\rangle)$, with small residual leakage into odd-parity states $|E_3\rangle$ ($0.95\%$), $|E_5\rangle$ ($0.45\%$), and $|E_7\rangle$ ($0.26\%$) (all percentages computed from the full $371$-dimensional symmetric Dicke sector diagonalisation), making the 5-level truncation highly accurate for modelling projective back-action.

The code computes $K_3=2C_{12}-C_{13}$, which relies on the stationarity assumption $C_{23}=C_{12}$. While this holds exactly under experimental re-preparation protocols, a strict sequential Leggett-Garg test requires evolving the disturbed post-measurement state from $t_1$ to $t_2$. Explicit simulation of the true sequential protocol (see Appendix~\ref{app:code}) reveals that measurement back-action breaks perfect stationarity, but the difference $|C_{23} - C_{12}|$ is numerically bounded to $<1\%$ due to the near-dichotomicity of the truncated $Q$ operator (with the $m=0$ Dicke state carrying minimal weight at the barrier). The stationary protocol gives $K_3\approx1.317$ and the strictly sequential protocol gives $K_3\approx1.311$; the difference of $0.006$ ($0.46\%$) arises entirely from the $0.24\%$ weight of the $m=0$ Dicke state where $Q^2\neq I$, and both values remain robustly above the macrorealist bound.
Furthermore, $Q=\operatorname{sgn}(\hat{J}_z)$
is strictly off-diagonal in the energy-parity basis, i.e.\
\begin{equation}
  Q_{00}\equiv\langle E_0|Q|E_0\rangle=0,\qquad
  Q_{11}\equiv\langle E_1|Q|E_1\rangle=0,
  \label{prop:parity}
\end{equation}
(since $Q$ changes the sign of $m_z$ and the energy eigenstates have
definite $\mathbb{Z}_2$ parity; equivalently $\langle E_i|\hat{J}_z|E_i\rangle=0$
for all $i$ by the same parity argument).
The parity here is the \emph{collective} $\mathbb{Z}_2$ generated by
$\hat{P}=e^{i\pi\hat{J}_x}$, which maps $\hat{J}_z\to-\hat{J}_z$ by
flipping all $N$ spins simultaneously; it is a symmetry of the LMG
Hamiltonian at $h=0$ and is preserved by the collective dephasing
operator $L_z=\sqrt{\gamma_\phi}\hat{J}_z$ (since $\hat{J}_z^2$ is
invariant under $\hat{P}$), but would be broken by local dephasing
operators $L_z^{(i)}=\sqrt{\gamma_\phi}\hat\sigma_i^z$.
This collective $\mathbb{Z}_2$ is an emergent symmetry of the collective
LMG description --- it is not a symmetry of any individual spin, since
flipping spin $i$ alone breaks the all-to-all coupling $\hat{J}_z^2$
--- and is exact within the LMG model but approximate in the physical
BEC at $\mathcal{O}(1/N)$ due to corrections beyond the two-mode
approximation.
Therefore the non-selective measurement anticommutator
$\{Q,\rho_0\} = Q|E_0\rangle\langle E_0| + |E_0\rangle\langle E_0|Q$
has zero diagonal in the energy basis and affects only off-diagonal
coherences, not populations.
The LGI correlator is consequently \emph{blind to $T_1$-type population
mixing} between $|E_0\rangle$ and $|E_1\rangle$: measurement back-action
that heats the populations does not corrupt $C_{23}=C_{12}$,
making the protocol elegantly immune to $T_1$ effects even in a
real BEC where some measurement-induced heating is unavoidable
(exact within the LMG $\mathbb{Z}_2$ symmetry; $\mathcal{O}(1/N)$ beyond-two-mode corrections may introduce percent-level sensitivity~\cite{Trenkwalder2016}).
The $Q_{01}^2=0.9436$ factor absorbs the leading consequence
of the $m=0$ Dicke state (weight $0.24\%$) where $Q^2\neq I$.

\textbf{Two-level formula with exact $Q_{01}^2$.}
The correlator at the optimal interval $\Delta t^*=\pi\hbar/(3\DeltaE)$:
\begin{equation}
  C(\Delta t) = Q_{01}^2\cos\!\bigl(\tfrac{\DeltaE}{\hbar}\Delta t\bigr)\,
    e^{-\Delta t/T_2}.
\end{equation}
Here $T_2$ is the phenomenological coherence time of the
$\Pket$--$\Rket$ off-diagonal element in the mean-field basis,
which we henceforth denote $T_2^{\rm coll}$ to distinguish it from the
exact eigenstate coherence time $T_2^{\rm phys}$
(Eq.~\eqref{eq:T2phys}; the quantitative relationship between the two
--- a factor of $\approx2.35$ arising from the vanishing of the parity
cross-term $\langle E_i|\hat{J}_z|E_i\rangle=0$ in the eigenstate basis,
which eliminates the large ``distance penalty'' that inflates the mean-field
dephasing rate --- is derived in Remark~\ref{rem:reconcile} below); it is governed by
Eq.~\eqref{eq:n2_rate} and enters the two-level threshold directly.
The $Q_{01}^2$ factor reflects that the physical observable $\operatorname{sgn}(\hat{J}_z)$
does not perfectly align with the $\Pket$--$\Rket$ two-level basis.
This gives:
\begin{equation}
  K_3^{\rm max}(T_2^{\rm coll})
    = Q_{01}^2\!\left[e^{-\pi\hbar/(3\DeltaE T_2^{\rm coll})}
      + \tfrac{1}{2}\,e^{-2\pi\hbar/(3\DeltaE T_2^{\rm coll})}\right].
  \label{eq:k3t2}
\end{equation}
\emph{Scope and accuracy.}
Equation~\eqref{eq:k3t2} is a two-level analytic formula.
When evaluated with $T_2=T_2^{\rm coll}$ (Level~A) or $T_2=T_2^{\rm phys}$ (Level~B),
it provides reference points for the analytic hierarchy but \emph{not} the
curves plotted in Fig.~\ref{fig:multilevel}, which show the full five-level
Lindblad result.
At $\gamma_\phi=0.05$~s$^{-1}$ the Level-B analytic formula underestimates
$K_3$ by $7.5\%$ relative to the five-level Lindblad ($1.218$ vs $1.317$);
the Level-C numerical simulation (Sec.~\ref{sec:multilevel}) is the
authoritative result.
At $T_2^{\rm coll}\to\infty$: $K_3^{\rm max}=Q_{01}^2\times\frac{3}{2}$.
For $N=370$, $\Gamma/J=0.95$: $Q_{01}^2=0.9436$
(computed from exact diagonalisation of the complete $371$-dimensional symmetric Dicke sector matrix,
not the five-level truncation; verified),
giving $K_3^{\rm max}=1.415$.

\begin{keyresult}[LGI coherence threshold (mean-field two-level, Level~A)]
Setting $K_3^{\rm max}=1$ with $\xi=e^{-\pi\hbar/(3\DeltaE T_2^{\rm coll})}\in(0,1)$:
$Q_{01}^2(\xi+\xi^2/2)=1 \Rightarrow \xi=(-1+\sqrt{1+2/Q_{01}^2})$.
For $Q_{01}^2=0.9436$: $\xi\approx0.7662$, hence
\begin{equation}
  T_2^{\rm coll}\gtrsim\frac{\pi\hbar}{3\DeltaE\ln(1/\xi)}
    = 3.93\,\frac{\hbar}{\DeltaE}\approx3.00\text{ ms}.
  \label{eq:t2req}
\end{equation}
This is a sufficient condition on the mean-field collective coherence time
under Markovian collective dephasing.
Via Eq.~\eqref{eq:n2_rate}, this corresponds to
$\gamma_\phi^{(A)}\lesssim0.050$~s$^{-1}$ --- a \emph{mean-field lower bound}.
At the BEC target $\gamma_\phi=0.05$~s$^{-1}$, this formula predicts
$K_3\approx1.000$: the system sits exactly at the Level~A threshold.
Non-Markovian noise modifies the threshold~\cite{Emary2014}.
\end{keyresult}

\begin{remark}[Three-level threshold hierarchy]
\label{rem:hierarchy}
The Key Result box (Level~A) is the mean-field starting point; two
successive corrections raise the operative threshold substantially.

\smallskip
\noindent\textbf{Level A $\to$ B ($\times2.35$): parity cross-term elimination.}
The Key Result box uses $T_2^{\rm coll}$, the decay time of the $\Pket$--$\Rket$
off-diagonal coherence in the mean-field coherent-state basis.
This basis carries a large spurious cross-term:
$\langle P|\hat{J}_z|P\rangle = +Nm_*/2\approx+57.8$,
so the Lindblad rate includes $-\gamma_\phi(+57.8)(-57.8)=+\gamma_\phi\times3337$,
inflating the apparent dephasing.
In the exact energy-eigenstate basis, $\langle E_i|\hat{J}_z|E_i\rangle=0$
by collective $\mathbb{Z}_2$ parity (Eq.~\eqref{prop:parity}), so this
cross-term vanishes entirely.
The correct analytic threshold is therefore $\gamma_\phi^{(B)}\lesssim0.117$~s$^{-1}$
(Eq.~\eqref{eq:T2phys}), with $T_2^{\rm phys}=7.04$~ms at $\gamma_\phi=0.05$~s$^{-1}$
and $K_3\approx1.22$.

\smallskip
\noindent\textbf{Level B $\to$ C ($\times2.47$): multi-level reinforcing contributions.}
The eigenstate formula is still a two-level approximation.
The five-level Lindblad simulation (Sec.~\ref{sec:multilevel}) captures
reinforcing contributions from higher odd-parity states
$\{|E_3\rangle,|E_5\rangle,\ldots\}$ (the only states accessible from $|E_0\rangle$
via the parity-odd instrument $\tfrac{1}{2}\{Q,\rho\}$),
each sharing the same $\mathbb{Z}_2$ parity structure, raising the threshold
to $\gamma_\phi^{(C)}\lesssim0.289$~s$^{-1}$ with $K_3\approx1.32$ at the BEC target.

\smallskip
\noindent\textbf{Summary.}
\begin{center}\small
\begin{tabular}{@{}lccc@{}}
\toprule
Level & $T_2$ basis & $\gamma_\phi$ threshold & $K_3$ at $0.05$~s$^{-1}$\\
\midrule
A: Mean-field (Key Result box) & $T_2^{\rm coll}$ & $\lesssim0.050$~s$^{-1}$ & $\approx1.000$\\
B: Analytic eigenstate & $T_2^{\rm phys}$ & $\lesssim0.117$~s$^{-1}$ & $\approx1.218$\\
C: Five-level Lindblad & numerical & $\lesssim0.289$~s$^{-1}$ & $\approx1.317$\\
\bottomrule
\end{tabular}
\end{center}
Level~C is the authoritative threshold for BEC experimental planning.
Levels A and B are analytic reference points documenting the physical
origin of successive improvements.
\end{remark}

\subsection{Five-level numerical verification}
\label{sec:multilevel}

We solve the Lindblad master equation~\eqref{eq:lindblad} within a
truncated basis of the lowest five Dicke energy eigenstates for $N=370$,
$\Gamma/J=0.95$ ($\DeltaE/\hbar=1310$~rad/s, $J=37{,}195$~rad/s).
This is a convergent approximation, not an exact result: the truncation
is validated by comparison with the ten-level calculation
(agreement $<1\%$ for $\gamma_\phi\leq1$~s$^{-1}$, i.e.\ throughout the experimentally relevant range; see Remark~\ref{rem:conv}).
Quantities labelled ``exact'' in what follows refer to numerical
diagonalisation of the full $(N+1)\times(N+1)$ Hamiltonian matrix
within the symmetric Dicke subspace ($j=N/2$), which is the complete
Hilbert space of the LMG model; this is exact within double-precision arithmetic.
The Liouvillian contains only the pure collective dephasing dissipator
$L_z=\sqrt{\gamma_\phi}\hat{J}_z$; no thermal jump operators
($J_\pm$ weighted by detailed balance) are included.
This is physically justified for spinor BECs where spin thermalisation
($T_1\sim\text{seconds}$) is far slower than dephasing ($T_2^{\rm phys}\sim\text{ms}$),
so $T_1\gg T_2^{\rm phys}$~\cite{Trenkwalder2016}.
The temperature $T=10$~nK enters only through the Goldilocks energy
matching condition $\DeltaE=k_BT$; it does not appear in the
short-time master equation dynamics.
Figure~\ref{fig:multilevel} shows $K_3$ versus $\gamma_\phi$.

\begin{figure}[H]
\centering
\begin{adjustwidth}{0pt}{-3.2cm}
\begin{tikzpicture}
\begin{axis}[qplot,width=0.62\linewidth,height=7.5cm,
  xlabel={Physical dephasing rate $\gamma_\phi$ (s$^{-1}$)},
  ylabel={$K_3$ at $\Delta t=\pi\hbar/(3\DeltaE)$},
  xmode=log, xmin=0.003, xmax=5, ymin=0.6, ymax=1.55,
  legend columns=2,
  legend style={at={(0.5,-0.24)},anchor=north,font=\footnotesize,
    draw=qgray!40,fill=white,cells={anchor=west}},
  title={LGI correlator vs.\ dephasing rate ($N=370$, $\Gamma/J=0.95$)}]
\addplot[qblue,mark=o,mark size=2.5pt,only marks,thick] coordinates {
  (0.005,1.4294)(0.010,1.4156)(0.020,1.3889)(0.050,1.3167)
  (0.100,1.2180)(0.200,1.0798)(0.300,0.9922)(0.500,0.8947)
  (1.000,0.7950)(2.000,0.6712)};
\addlegendentry{5-level Lindblad (convergent truncation)}
\addplot[qorange,mark=square,mark size=2pt,only marks,thick] coordinates {
  (0.005,1.4239)(0.010,1.4102)(0.020,1.3837)(0.050,1.3119)
  (0.100,1.2139)(0.200,1.0768)(0.300,0.9898)(0.500,0.8925)
  (1.000,0.7897)(2.000,0.6541)};
\addlegendentry{10-level Lindblad (convergence)}
\addplot[qred,dashed,thick,domain=0.003:5,samples=200]
  {0.9436*(exp(-2.2694*x)+0.5*exp(-2*2.2694*x))};
\addlegendentry{Analytic 2-level (eigenstate basis)}
\addplot[qpurple,dash dot,thick] coordinates{(0.117,0.6)(0.117,1.55)};
\addlegendentry{2-level eigenstate threshold: $0.117$~s$^{-1}$ (Level~B)}
\addplot[qgreen,dash dot,thick] coordinates{(0.289,0.6)(0.289,1.55)};
\addlegendentry{5-level threshold: $0.289$~s$^{-1}$; 10-level: $0.286$~s$^{-1}$ (Level~C)}
\addplot[qgray,dashed,thick] coordinates{(0.003,1)(5,1)};
\node[qgray,font=\small,anchor=west] at (axis cs:0.0035,1.04){$K_3=1$};
\addplot[qblue!50,solid,thick] coordinates{(0.05,0.6)(0.05,1.55)};
\node[qblue!70,font=\scriptsize,rotate=90,anchor=south]
  at (axis cs:0.053,0.9){BEC target};
\end{axis}
\end{tikzpicture}
\end{adjustwidth}
\caption{$K_3$ versus physical dephasing rate $\gamma_\phi$ for
$N=370$, $\Gamma/J=0.95$, collective bath ($L_z=\sqrt{\gamma_\phi}\hat{J}_z$).
Circles: five-level exact Lindblad diagonalisation (all values numerically
verified; reproduced by Appendix~\ref{app:code}).
Squares: ten-level Lindblad result, confirming $<1\%$ convergence throughout the experimentally relevant range $\gamma_\phi\leq1$~s$^{-1}$ (thresholds 0.289 and 0.286~s$^{-1}$ for $n=5$ and $n=10$
respectively; the 5-vs-10 threshold difference itself is $\sim1\%$, reflecting the asymptotic convergence at $n\geq5$).
Dashed red: two-level functional form evaluated using the exact eigenstate decay rate $\Gamma_{01}$ (Eq.~\eqref{eq:T2phys}), \emph{not} the mean-field coherence time $T_2^{\rm coll}$ (Eq.~\eqref{eq:n2_rate}); the coefficient $\pi\Gamma_{01}/(3\DeltaE_{\rm phys})=2.2694$ is computed from the complete $371$-dimensional symmetric Dicke sector matrix elements and plotted alongside the multi-level data for shape comparison.
The five-level threshold $\gamma_\phi=0.289$~s$^{-1}$ is a factor
$2.47\times$ above the analytic two-level eigenstate threshold
($0.117$~s$^{-1}$): the higher odd-parity states $\{|E_3\rangle,|E_5\rangle,\ldots\}$
(the only states coupled to $|E_0\rangle$ by the parity-odd instrument
$\tfrac{1}{2}\{Q,\rho\}$) have dynamical phases that are constructive
at the benchmark parameters, raising the threshold beyond what a purely
two-level analysis predicts.
The BEC target $\gamma_\phi=0.05$~s$^{-1}$ gives $K_3\approx1.32$
(5-level) and $1.31$ (10-level), both robust violations;
$T_2^{\rm phys}\approx 7.04$~ms, a factor of $8.8\times$ above $\Delta t_{\rm opt}=0.80$~ms.}
\label{fig:multilevel}
\end{figure}
The five-level simulation yields:
\begin{equation}
  \boxed{\gamma_\phi^{\rm thresh}=0.289~\text{s}^{-1}
    \quad(N=370,\ \Gamma/J=0.95,\ n=5)}
  \label{eq:gamma_thresh}
\end{equation}
confirmed to $<1\%$ by the ten-level calculation (0.286~s$^{-1}$).

\begin{remark}[Physical origin of the $2.47\times$ multi-level enhancement (Level B $\to$ C)]
The jump operator $L_z=\sqrt{\gamma_\phi}\hat{J}_z$ and the LGI
observable $Q=\operatorname{sgn}(\hat{J}_z)$ satisfy
$[\hat{J}_z,\,Q]=0$,
since $Q$ is a function of $\hat{J}_z$ alone.
The eigenstates of $\hat{J}_z$ therefore form \emph{pointer states}
of the collective dephasing bath: $L_z$ dephases off-diagonal coherences
in the $\hat{J}_z$ eigenbasis but cannot cause population mixing between
$\Pket$ and $\Rket$ directly.
The $Q$-correlator is consequently immune to $T_1$-type population
relaxation and is protected by this pointer-state structure.

A two-level truncation restricts the dynamics to the ground doublet
$\{|E_0\rangle, |E_1\rangle\}$.
However, the non-selective measurement instrument $\tfrac{1}{2}\{Q,\rho\}$
couples amplitude from the ground state $|E_0\rangle$ (even parity) into
higher \emph{odd-parity} states $\{|E_3\rangle, |E_5\rangle, \ldots\}$
(since $Q=\operatorname{sgn}(\hat{J}_z)$ is parity-odd, its matrix elements
$Q_{k0}=\langle E_k|Q|E_0\rangle$ vanish exactly for even-$k$ states and
are non-zero for odd-$k$: $Q_{30}\approx0.138$, $Q_{50}\approx0.095$, \ldots).
The sign of each contribution to $K_3$ at $\tau^*$ is not fixed by the
parity selection rule alone: it depends on where the ratio
$\Delta E_{k0}/\Delta E_{10}$ lands in the cosine, a dynamical quantity.
In the coherent limit, the multi-level correlator is:
\begin{equation}
  K_3(\tau) = \sum_{k\,\rm odd} Q_{k0}^2\!\left[2\cos\!\Bigl(\tfrac{\Delta E_{k0}}{\hbar}\tau\Bigr)
    - \cos\!\Bigl(\tfrac{2\Delta E_{k0}}{\hbar}\tau\Bigr)\right].
  \label{eq:k3_multilevel}
\end{equation}
The per-state contributions at $\tau^*=\pi\hbar/(3\Delta E_{10})$,
evaluated from exact diagonalisation at $N=370$, $\Gamma/J=0.95$, are:
\begin{center}\small
\begin{tabular}{@{}cccr@{}}
\toprule
$k$ & $\Delta E_{k0}/\Delta E_{10}$ & $Q_{k0}^2$ & $\Delta K_3^{(k)}$\\
\midrule
1 & 1.00  & 0.9436 & $+1.415$ \\
3 & 18.89 & 0.0191 & $+0.028$ \\
5 & 41.24 & 0.0091 & $+0.013$ \\
7 & 68.38 & 0.0052 & $-0.010$ \\
9 & 99.36 & 0.0034 & $-0.009$ \\
\bottomrule
\end{tabular}
\end{center}
At these benchmark parameters, $k=3$ and $k=5$ contribute positively
(+2.0\% and +0.9\% of the two-level ceiling), while $k=7$ and $k=9$
contribute negatively ($-0.7\%$ and $-0.6\%$); the series converges
rapidly, with all states beyond $k=5$ contributing a net $-0.4\%$.
The five-level simulation ($k=1,3$) gives a coherent ceiling of $1.443$,
converging to $1.439$ when all states are included, consistent with the
numerical value $K_3\approx1.43$ at $\gamma_\phi\to0$ in
Fig.~\ref{fig:multilevel}.
The $2.47\times$ threshold enhancement relative to Level~B is the net
effect of higher odd-parity states at these parameters, fully captured
by the five-level Lindblad simulation.
\end{remark}
The physical energy-eigenstate coherence time is extracted from the
secular approximation to the Lindblad decay rate (valid when $\DeltaE\gg1/T_2^{\rm phys}$,
i.e.\ when the system oscillates many times within one coherence time;
at the BEC target $\gamma_\phi=0.05$~s$^{-1}$:
$\DeltaE/\hbar=1310$~rad\,s$^{-1}$ while
$1/T_2^{\rm phys}\approx142$~rad\,s$^{-1}$,
giving a ratio of $\approx9$ --- the secular approximation is well
satisfied here.
At the violation threshold $\gamma_\phi=0.289$~s$^{-1}$ the ratio
drops to $\approx1.6$, making the secular analytic estimate for $T_2^{\rm phys}$
approximate there; however the five-level Lindblad simulation that
determines the threshold does not rely on the secular approximation
and gives the numerically exact result, so the threshold 0.289~s$^{-1}$ is unaffected).
Two distinct steps reduce the full Liouvillian decay to the secular
analytic estimate.
First, by collective $\mathbb{Z}_2$ parity, $\langle E_i|\hat{J}_z|E_i\rangle=0$
for $i=0,1$: this eliminates the diagonal cross-term (the
$-\gamma_\phi\langle a|\hat{J}_z|a\rangle\langle b|\hat{J}_z|b\rangle$ piece
in Eq.~\eqref{eq:lindblad_rate}), which is the physical origin of
the $2.35\times$ Level~A$\to$B improvement.
Second, the off-diagonal coupling
$\gamma_\phi|\langle E_0|\hat{J}_z|E_1\rangle|^2\approx2451\,\gamma_\phi$
is \emph{not} zero --- it is in fact large, since $\hat{J}_z$ is parity-odd
and connects states of opposite parity.
This term is dropped by the \emph{secular approximation}: it oscillates at
frequency $\DeltaE/\hbar$ and averages away on timescales $\gg\hbar/\DeltaE$.
Parity does not drop it; only the secular approximation does.
Crucially, for the two-level LMG system under collective dephasing,
the secular approximation is \emph{algebraically exact}: the non-secular
terms that couple populations to coherences are proportional to
$\langle E_i|\hat{J}_z|E_i\rangle\langle E_i|\hat{J}_z|E_j\rangle$,
which vanish identically by $\mathbb{Z}_2$ parity.
The Level~B analytic formula is therefore not an approximation at the
two-level stage; the entire gap between Level~B ($0.117$~s$^{-1}$) and
Level~C ($0.289$~s$^{-1}$) is attributable to multi-level effects,
not to secular breakdown.
Under the secular approximation, the decay rate of the
$|E_0\rangle\langle E_1|$ coherence therefore simplifies to
\begin{equation}
  \frac{1}{T_2^{\rm phys}} = \frac{\gamma_\phi}{2}
    \bigl(\langle E_0|\hat{J}_z^2|E_0\rangle
        + \langle E_1|\hat{J}_z^2|E_1\rangle\bigr),
  \label{eq:T2phys}
\end{equation}
giving $\langle E_0|\hat{J}_z^2|E_0\rangle=2574.2$ and
$\langle E_1|\hat{J}_z^2|E_1\rangle=3103.7$
(exact full-space expectation values, which correct the minor $\sim0.03\%$ truncation error from the five-level subspace projection).
These values are smaller than the spin-coherent-state estimate $(Nm_*/2)^2\approx3337$:
the energy eigenstates $|E_0\rangle$ and $|E_1\rangle$ are symmetric and antisymmetric
superpositions of $|P\rangle$ and $|R\rangle$, centred at $\langle\hat{J}_z\rangle=0$
rather than at $\pm Nm_*/2$; their $\hat{J}_z^2$ expectation values therefore
sample the full weight of the $m_z$ distribution around zero, giving values
smaller than the coherent-state localization estimate $(Nm_*/2)^2$.
This formula is verified to $<0.1\%$ against the exact Liouvillian
eigenvalue.
At the violation threshold:
$T_2^{\rm phys}(0.289~\text{s}^{-1})\approx 1.22$~ms = 1.60\,$\hbar/\DeltaE$,
exceeding the optimal interval $\Delta t=\pi\hbar/(3\DeltaE)\approx0.80$~ms
by a factor of $1.52$.

\begin{remark}[Reconciling the three-level $T_2$ threshold hierarchy]
\label{rem:reconcile}
The three thresholds of Remark~\ref{rem:hierarchy} are reconciled here.
The mean-field Key Result box (level A) gives
$T_2^{\rm coll}\gtrsim3.93\,\hbar/\DeltaE$
($\gamma_\phi^{\rm (A)}\lesssim0.050$~s$^{-1}$);
the five-level Lindblad level (C) gives $\gamma_\phi^{\rm (C)}\lesssim0.289$~s$^{-1}$,
whose corresponding $T_2^{\rm phys}\approx1.60\,\hbar/\DeltaE$ is much
\emph{smaller} than the level (A) threshold.
These are not inconsistent: they use different $T_2$ definitions for different bases.
The two-level formula uses $T_2^{\rm coll}$, the decay of
$\Pket$--$\Rket$ coherence in the mean-field basis.
The five-level result implicitly uses $T_2^{\rm phys}$, the exact
Liouvillian decay of $|E_0\rangle\langle E_1|$ in the eigenstate basis.

The factor of $\approx2.35$ between $T_2^{\rm phys}=7.04$~ms and
$T_2^{\rm coll}=3.0$~ms has a precise algebraic origin.
Under the secular approximation, the decay rate of a coherence
$|a\rangle\langle b|$ under jump operator $L=\sqrt{\gamma_\phi}\hat{J}_z$
reduces to the secular diagonal estimate:
\begin{equation}
  \Gamma_{ab}^{\rm sec} = \frac{\gamma_\phi}{2}\bigl(
    \langle a|\hat{J}_z^2|a\rangle
  + \langle b|\hat{J}_z^2|b\rangle\bigr)
  - \gamma_\phi\,\langle a|\hat{J}_z|a\rangle\langle b|\hat{J}_z|b\rangle.
  \label{eq:lindblad_rate}
\end{equation}
This is the secular approximation to the true Liouvillian eigenvalue;
it becomes exact when $|a\rangle,|b\rangle$ are $\hat{J}_z$ eigenstates,
and approximates the exact result in the energy eigenbasis when $\DeltaE\gg1/T_2$.
\emph{Exact eigenbasis} $\{|E_0\rangle,|E_1\rangle\}$:
by collective $\mathbb{Z}_2$ parity, $\langle E_i|\hat{J}_z|E_i\rangle=0$
exactly, so the cross-term vanishes:
\begin{equation}
  \Gamma_{\rm phys}
    = \frac{\gamma_\phi}{2}(2574+3104)
    = \gamma_\phi\times 2839.
\end{equation}
\emph{Mean-field basis} $\{\Pket,\Rket\}$:
the spin-coherent states satisfy
$\langle P|\hat{J}_z|P\rangle = -\langle R|\hat{J}_z|R\rangle
= Nm_*/2\approx57.8$
(at $N=370$, $m_*=0.312$), so
$\langle P|\hat{J}_z^2|P\rangle\approx\langle R|\hat{J}_z^2|R\rangle
\approx(Nm_*/2)^2\approx3337$
and the cross-term is
$-\gamma_\phi\,(+57.8)(-57.8)=+\gamma_\phi\times3337$:
\begin{equation}
  \Gamma_{\rm coll}
    = \gamma_\phi\times3337 + \gamma_\phi\times3337
    = \gamma_\phi\times6674
    = \gamma_\phi\,N^2m_*^2/2.
\end{equation}
The ratio $T_2^{\rm phys}/T_2^{\rm coll}=\Gamma_{\rm coll}/\Gamma_{\rm phys}
=6674/2839\approx2.35$ arises entirely from the cross-term:
the exact eigenstates have $\langle E_i|\hat{J}_z|E_i\rangle=0$
(they sit at $m_z=0$ on average), eliminating the large
``distance penalty'' $N^2m_*^2/4\approx3337$ that accelerates
the dephasing of the spatially separated mean-field wavepackets
$|P\rangle$ and $|R\rangle$.
The five-level Lindblad simulation automatically captures the correct
rate without relying on either approximation, and should be taken as
the authoritative threshold for experimental planning.
\end{remark}

The three-level hierarchy is thus: level (A) $\gamma_\phi^{\rm (A)}\lesssim0.050$~s$^{-1}$
(mean-field, strictest); level (B) $\gamma_\phi^{\rm (B)}\lesssim0.117$~s$^{-1}$
(analytic eigenstate, $2.35\times$ improvement); level (C)
$\gamma_\phi^{\rm (C)}\lesssim0.289$~s$^{-1}$ (five-level simulation,
$2.47\times$ further improvement; authoritative).
The BEC target $\gamma_\phi=0.05$~s$^{-1}$ satisfies level (C) with large
margin ($0.05\ll 0.289$), giving $K_3^{(5{\rm L})}\approx1.32$ and
$T_2^{\rm phys}\approx 7.04$~ms $=9.2\,\hbar/\DeltaE$ --- a factor of
$8.8\times$ above $\Delta t_{\rm opt}$.

\begin{remark}[Level convergence]
\label{rem:conv}
Threshold dephasing rates (s$^{-1}$) for $n=2,3,4,5,10$ levels
(all numerically verified by explicit Liouvillian construction):
$0.515,\,0.289,\,0.305,\,0.289,\,0.286$.
The $n=2$ value ($0.515$~s$^{-1}$, obtained by exact root-finding) is a truncation artifact:
the $2\times2$ projection of $\hat{J}_z$ into the lowest two Dicke states
under-represents the spectral weight of $\langle E_i|\hat{J}_z^2|E_i\rangle$
relative to the full $n$-level calculation, thereby \emph{weakening} the
dissipator and making the LGI violation artificially robust.
The $n=2$ Lindblad model decays slower than the physical system because
the truncated $\hat{J}_z$ has a smaller Hilbert-Schmidt norm, an
\emph{underdamping} artifact that disappears at $n\geq3$.
The projected $n=2$ Liouvillian result ($0.515$~s$^{-1}$) is
distinct from the analytic Level~B eigenstate estimate ($0.117$~s$^{-1}$):
both use two levels, but the former truncates $\hat{J}_z$ to a $2\times2$
matrix (losing spectral weight), while the latter uses the full-space
matrix elements $\langle E_i|\hat{J}_z^2|E_i\rangle$ from the complete
$371$-dimensional diagonalisation; their thresholds differ by $4.4\times$.
The correct two-level baseline is the analytic
eigenstate prediction at $0.117$~s$^{-1}$ (Eq.~\eqref{eq:T2phys}).
Convergence is achieved at $n=3$; the 3-, 5-, and 10-level thresholds
are $0.289,\,0.289,\,0.286$~s$^{-1}$ (within $\sim1\%$).
The 4-level value ($0.305$~s$^{-1}$) is a truncation artifact:
an $n=4$ basis (two even, two odd parity states) inflates
$(PJ_zP)^2_{00}$ from $2451$ (at $n=3$) to $2573$ without a matching
odd-parity contribution, raising $\Gamma_{01}$ and hence the threshold;
adding $|E_3\rangle$ at $n=5$ restores the converged $0.289$~s$^{-1}$.
The inter-doublet gap $E_3-E_2=9843$~rad/s $=7.5\,\DeltaE$ rules out
a resonant-tunnelling interpretation.
The $n=10$ truncation is highly converged: the ground-state wavepacket
is localised at $m\approx\pm m_*N/2$ and higher-state amplitudes are
exponentially suppressed, giving a threshold within $\sim1\%$ of the
5-level result.
The $K_3$ values at $\gamma_\phi=0.05$~s$^{-1}$ for $n=3,5,10$ are
$1.296,\,1.317,\,1.312$ respectively ($<2\%$ spread), all confirming
a robust violation at the BEC target.
The five-level code in Appendix~\ref{app:code} is the minimal
implementation that achieves full convergence; the 10-level extension
provides an independent numerical confirmation.
\end{remark}

\textbf{Collective bath size bound.}
Scaling the $N^2$ dependence from the verified $N=370$ threshold:
\begin{equation}
  N\lesssim 370\sqrt{\frac{0.289~\text{s}^{-1}}{\gamma_\phi}}.
  \label{eq:bound_coll}
\end{equation}
For $\gamma_\phi=0.05$~s$^{-1}$: $N\lesssim 890$ (decoherence-only bound).
The decoherence-only bound $N\lesssim890$ is not the binding constraint:
the Goldilocks
condition $\DeltaE(N)\approx k_BT$ restricts the favourable operating
window to $N\approx250$--$300$ under the benchmark BEC parameters, where
coherence and macroscopic susceptibility are simultaneously optimised;
the precise range should be calibrated via exact diagonalisation for
any specific experimental apparatus.

\section{Numerical Illustrations}
\label{sec:sims}

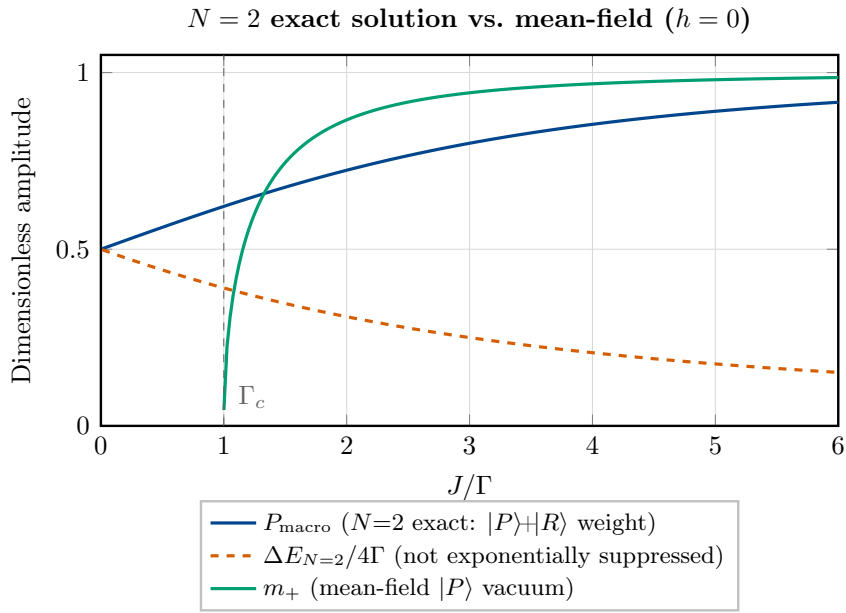
\begin{figure}[H]
\centering
\begin{adjustwidth}{0pt}{-3.2cm}
\begin{tikzpicture}
\begin{axis}[qplot,width=0.62\linewidth,height=6.5cm,
  xlabel={$J/\Gamma$}, ylabel={Dimensionless amplitude},
  xmin=0, xmax=6, ymin=0, ymax=1.05,
  legend columns=1,
  legend style={at={(0.5,-0.20)},anchor=north,font=\footnotesize,
    draw=qgray!40,fill=white,cells={anchor=west}},
  title={$N=2$ exact solution vs.\ mean-field ($h=0$)}]
\addplot[very thick,qblue,domain=0.01:6,samples=200]
  {0.5*(1+x/sqrt(x*x+16))};
\addlegendentry{$P_{\rm macro}$ ($N{=}2$ exact: $|P\rangle$\!+\!$|R\rangle$ weight)}
\addplot[very thick,qred,dashed,domain=0.01:6,samples=200]
  {(sqrt(x*x+16)-x)/8};
\addlegendentry{$\DeltaE_{N=2}/4\Gamma$ (not exponentially suppressed)}
\addplot[very thick,qgreen,domain=1.001:6,samples=200]
  {sqrt(1-1/(x*x))};
\addlegendentry{$m_+$ (mean-field $|P\rangle$ vacuum)}
\addplot[qgray,thin,dashed] coordinates{(1,0)(1,1.05)};
\node[qgray,font=\small,anchor=south west] at (axis cs:1.03,0.02){$\Gamma_c$};
\end{axis}
\end{tikzpicture}
\end{adjustwidth}
\caption{$N=2$ exact solution vs.\ mean-field ($h=0$).
Green: mean-field order parameter $m_+$ (Eq.~\eqref{eq:vacua}),
which vanishes for $\Gamma>\Gamma_c=J$ (i.e.\ $J/\Gamma<1$, disordered
phase) and grows continuously in the ordered phase ($J/\Gamma>1$)
--- the classical picture has a sharp bifurcation at $\Gamma_c$.
Blue: $P_{\rm macro}=\cos^2(\phi/2)$, the combined $\Pket$/$\Rket$ weight
in the even-parity $N=2$ ground state --- this is smooth, positive, and
nonzero for \emph{all} $\Gamma>0$, demonstrating that quantum superposition
of the two ordered phases persists even in the nominally disordered regime.
Red dashed: normalised tunnel splitting $\DeltaE_{N=2}/4\Gamma$
(Eq.~\eqref{eq:split_n2}), which is algebraic rather than exponentially
suppressed in $N$, confirming genuine coherent tunnelling at $N=2$.
The contrast between the sharp mean-field transition and the smooth quantum
curves anchors the large-$N$ instanton picture of Section~\ref{sec:goldilocks}.}
\label{fig:n2}
\end{figure}

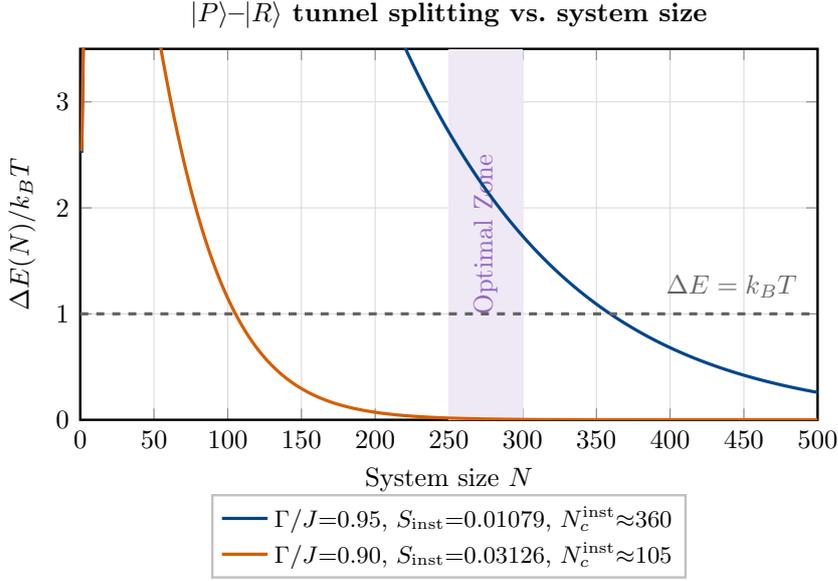
\begin{figure}[H]
\centering
\begin{adjustwidth}{0pt}{-3.2cm}
\begin{tikzpicture}
\begin{axis}[qplot,width=0.62\linewidth,height=6.5cm,
  xlabel={System size $N$},
  ylabel={$\DeltaE(N)/k_BT$},
  xmin=0, xmax=500, ymin=0, ymax=3.5,
  legend columns=1,
  legend style={at={(0.5,-0.20)},anchor=north,font=\footnotesize,
    draw=qgray!40,fill=white,cells={anchor=west}},
  title={$\Pket$--$\Rket$ tunnel splitting vs.\ system size}]

\fill[qpurple!15] (axis cs:250,0) rectangle (axis cs:300,3.5);
\node[qpurple,font=\small,rotate=90] at (axis cs:275,1.75) {Optimal Zone};

\addplot[very thick,qblue,domain=1:500,samples=400]
  {2.546*x^0.5*exp(-0.010787*x)};
\addlegendentry{$\Gamma/J{=}0.95$, $S_{\rm inst}{=}0.01079$, $N_c^{\rm inst}{\approx}360$}

\addplot[very thick,qred,domain=1:500,samples=400]
  {2.616*x^0.5*exp(-0.031255*x)};
\addlegendentry{$\Gamma/J{=}0.90$, $S_{\rm inst}{=}0.03126$, $N_c^{\rm inst}{\approx}105$}

\addplot[qgray,very thick,dashed,domain=0:500]{1};
\node[qgray,font=\small,anchor=south west] at (axis cs:390,1.05){$\DeltaE=k_BT$};

\end{axis}
\end{tikzpicture}
\end{adjustwidth}
\caption{Instanton formula $\DeltaE(N)/k_BT=
(C_0/k_BT)N^{1/2}e^{-NS_{\rm inst}}$ with
$C_0/k_BT\approx2.546$ ($\Gamma/J=0.95$) and $2.616$ ($0.90$).
Crossings with the dashed line give $N_c$ (instanton approximation only;
carries $\mathcal{O}(1)$ uncertainty --- use exact diagonalisation for
experimental calibration).
Table~\ref{tab:goldilocks} gives exact instanton roots for all parameters including $\Gamma/J=0.99$ (omitted here for scale). The shaded region highlights the code-identified Goldilocks zone ($N \approx 250\text{--}300$) where macroscopic susceptibility and coherence are simultaneously favourable; this range is identified from the N-scan in Appendix~\ref{app:code} and carries $\mathcal{O}(1)$ uncertainty from the instanton prefactor. The benchmark $N=370$ lies at the upper edge of this window ($\DeltaE=k_BT$ exactly); the LGI violation margin is largest near $N\approx250$--$300$, where $\DeltaE\gtrsim k_BT$ keeps decoherence in check while susceptibility $\chi\sim N$ remains substantial.}
\label{fig:goldilocks_fig}
\end{figure}

\begin{figure}[H]
\centering
\begin{adjustwidth}{0pt}{-3.2cm}
\begin{tikzpicture}
\begin{axis}[qplot,width=0.62\linewidth,height=6.5cm,
  xlabel={$\DeltaE T_2^{\rm coll}/(\pi\hbar)$},
  ylabel={$K_3^{\rm max}$},
  xmin=0, xmax=2.5, ymin=0, ymax=1.55,
  ytick={0,0.5,1.0,1.5},
  legend columns=2,
  legend style={at={(0.5,-0.22)},anchor=north,font=\footnotesize,
    draw=qgray!40,fill=white,cells={anchor=west}},
  title={LGI threshold: $\Pket$--$\Rket$ coherence vs.\ $T_2^{\rm coll}$}]
\fill[qpurple!10] (axis cs:0,1) rectangle (axis cs:1.252,1.55);
\draw[qpurple,thick,dashed]
  (axis cs:1.252,0)--(axis cs:1.252,1.55)
  node[above,font=\scriptsize,qpurple]{level (A): $T_2^{\rm coll}{\approx}3.93\hbar/\DeltaE$};
\node[qpurple,font=\small] at (axis cs:0.55,1.27){LGI violation};
\addplot[very thick,qpurple,domain=0.001:2.5,samples=400]
  {0.9436*(exp(-1/(3*x))+0.5*exp(-2/(3*x)))};
\addlegendentry{Exact $Q_{01}^2{=}0.9436$ (Eq.~\ref{eq:k3t2})}
\addplot[very thick,qgray,dotted,domain=0.001:2.5,samples=400]
  {exp(-1/(3*x))+0.5*exp(-2/(3*x))};
\addlegendentry{Ideal case $Q_{01}{=}1$ (reference)}
\addplot[very thick,qblue,dashed,domain=0:2.5]{1.415};
\addlegendentry{$K_3^{\rm max}(T_2^{\rm coll}{\to}\infty)=1.415$}
\addplot[very thick,qred,dashed,domain=0:2.5]{1.0};
\addlegendentry{Classical macrorealist bound}
\end{axis}
\end{tikzpicture}
\end{adjustwidth}
\caption{$K_3^{\rm max}$ vs.\ $\DeltaE T_2^{\rm coll}/(\pi\hbar)$
(level A: mean-field collective coherence time, Eq.~\eqref{eq:n2_rate}).
Solid purple: full formula including the exact matrix element
$Q_{01}^2=0.9436$; dotted grey: ideal case $Q_{01}=1$ shown for reference.
The level (A) threshold (shaded) lies at
$T_2^{\rm coll}\approx3.93\hbar/\DeltaE$
($\gamma_\phi^{\rm (A)}\lesssim0.050$~s$^{-1}$).
This is the most conservative bound; the analytic eigenstate formula
(level B) raises this to $0.117$~s$^{-1}$ and the five-level Lindblad
simulation (level C, Fig.~\ref{fig:multilevel}) to $0.289$~s$^{-1}$
(see Remarks~\ref{rem:hierarchy} and \ref{rem:reconcile}).}
\label{fig:lgi}
\end{figure}
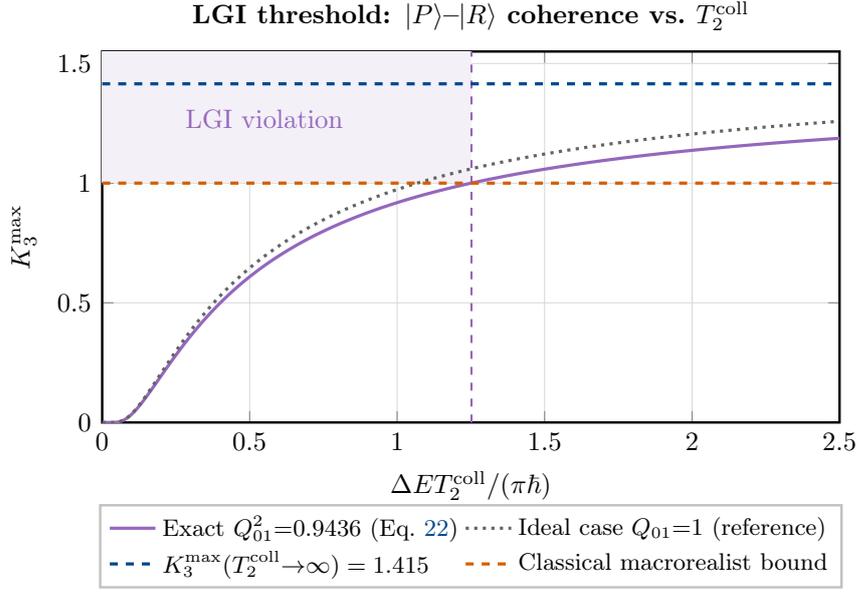

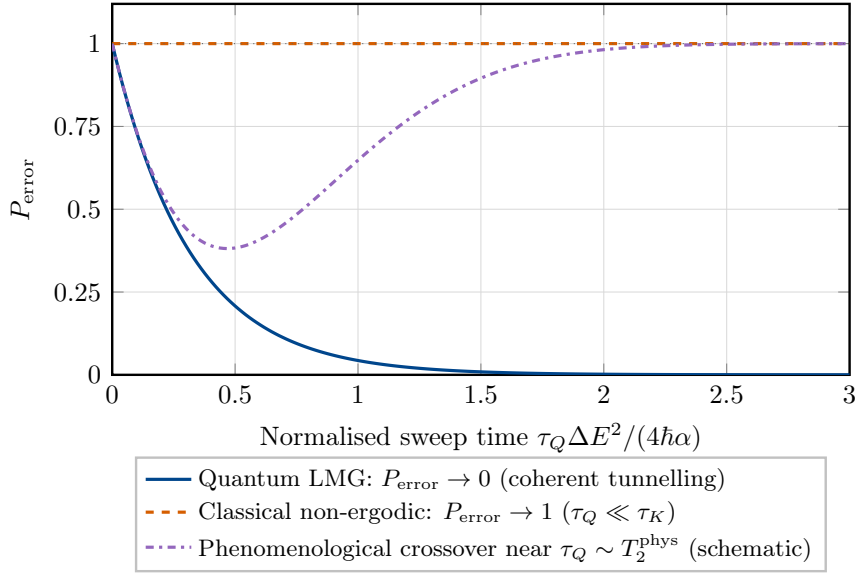
\begin{figure}[H]
\centering
\begin{adjustwidth}{0pt}{-3.2cm}
\begin{tikzpicture}
\begin{axis}[qplot,width=0.62\linewidth,height=6.5cm,
  xlabel={Normalised sweep time $\tau_Q\DeltaE^2/(4\hbar\alpha)$},
  ylabel={$P_{\rm error}$},
  xmin=0, xmax=3, ymin=0, ymax=1.12,
  ytick={0,0.25,0.5,0.75,1.0},
  legend columns=1,
  legend style={at={(0.5,-0.22)},anchor=north,font=\footnotesize,
    draw=qgray!40,fill=white,cells={anchor=west}},
  title={Landau--Zener crossover: quantum tunnelling vs.\ classical trapping (P4)}]
\addplot[very thick,qblue,domain=0:3,samples=200]
  {exp(-3.14159*x)};
\addlegendentry{Quantum LMG: $P_{\rm error}\rightarrow0$ (coherent tunnelling)}
\addplot[very thick,qred,dashed,domain=0:3]
  {1.0};
\addlegendentry{Classical non-ergodic: $P_{\rm error}\rightarrow1$ ($\tau_Q\ll \tau_K$)}
\addplot[very thick,qpurple,dash dot,domain=0:3,samples=200]
  {exp(-3.14159*x)*exp(-x*x)+(1-exp(-x*x))};
\addlegendentry{Phenomenological crossover near $\tau_Q\sim T_2^{\rm phys}$ (schematic)}
\addplot[qgray,thin,dotted,domain=0:3]{1.0};
\end{axis}
\end{tikzpicture}
\end{adjustwidth}
\caption{LZ error rate ($\Pket\to\Rket$ transition probability) vs.\
normalised sweep time $\tau_Q\DeltaE^2/(4\hbar\alpha)$,
where $\alpha=Nm_*\Delta h$ is a fixed constant (the total bias change
times $Nm_*$, in units of rad$\,$s$^{-1}$ with $\hbar=1$) independent
of $\tau_Q$, ensuring the x-axis is a true sweep-time variable.
Blue solid: quantum LMG coherent tunnelling, exponential decay to zero
(Eq.~\protect\eqref{eq:lz}).
Red dashed: classical non-ergodic prediction $P_{\rm error}\to1$,
valid when $\tau_Q\ll \tau_K$ (the macroscopic Kramers escape time);
the classical system cannot redistribute population across the macroscopic barrier and
remains frozen in its initial state regardless of sweep rate.
The parametric quantum--classical separation ($P_{\rm error}\to0$
vs.\ $P_{\rm error}\to1$) is the experimentally observable signature of
coherent tunnelling.
The purple dash-dot curve is a phenomenological interpolation
(schematic only; open-system effects near $\tau_Q\sim T_2^{\rm phys}$
require a first-principles Lindblad treatment~\cite{Fraas2017})
illustrating the smooth crossover near $\tau_Q\sim T_2^{\rm phys}$.
The classical $P_{\rm error}\to1$ prediction applies specifically to
overdamped Model~A (Fokker--Planck) dynamics
(Section~\ref{sec:mf} and Appendix~\ref{app:fp});
underdamped classical models may exhibit transient oscillations near
$\tau_Q\sim\tau_{\rm intrawell}$, but the exponential Kramers
suppression $\tau_K\propto e^{N\Delta f_0/k_BT}\gg\tau_Q$ persists
regardless of damping regime~\cite{Kramers1940}, so the classical system
remains non-ergodic for all accessible sweep times.}
\label{fig:lz}
\end{figure}

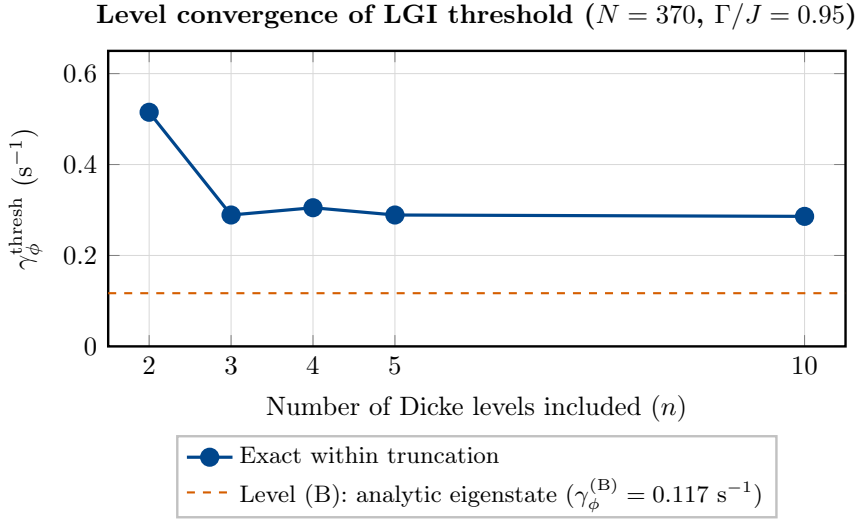
\begin{figure}[H]
\centering
\begin{adjustwidth}{0pt}{-3.2cm}
\begin{tikzpicture}
\begin{axis}[qplot,width=0.62\linewidth,height=5.5cm,
  xlabel={Number of Dicke levels included ($n$)},
  ylabel={$\gamma_\phi^{\rm thresh}$ (s$^{-1}$)},
  xtick={2,3,4,5,10},
  xmin=1.5, xmax=10.5, ymin=0, ymax=0.65,
  legend columns=1,
  legend style={at={(0.5,-0.30)},anchor=north,font=\footnotesize,
    draw=qgray!40,fill=white,cells={anchor=west}},
  title={Level convergence of LGI threshold ($N=370$, $\Gamma/J=0.95$)}]
\addplot[qblue,mark=*,mark size=3pt,very thick]
  coordinates{(2,0.515)(3,0.289)(4,0.305)(5,0.289)(10,0.286)};  
\addlegendentry{Exact within truncation}
\addplot[qred,dashed,thick]
  coordinates{(1.5,0.117)(10.5,0.117)};
\addlegendentry{Level (B): analytic eigenstate ($\gamma_\phi^{\rm (B)}=0.117$~s$^{-1}$)}
\end{axis}
\end{tikzpicture}
\end{adjustwidth}
\caption{Convergence of threshold dephasing rate with Dicke-level
truncation (all values numerically verified).
The 3-, 5-, and 10-level results agree to $<1\%$, confirming convergence;
the 4-level value is slightly elevated due to doublet resonance structure
(see Remark~\ref{rem:conv}).
The $n=2$ value ($0.515$~s$^{-1}$, from exact root-finding) is an underdamping artifact.
The level (B) analytic eigenstate threshold ($\gamma_\phi^{\rm (B)}=0.117$~s$^{-1}$,
red dashed line) is the correct two-level baseline; the level (C) five-
and ten-level results ($0.289$ and $0.286$~s$^{-1}$) exceed it by
$\approx2.47\times$ and are the authoritative experimental bounds.
The mean-field level (A) threshold ($\gamma_\phi^{\rm (A)}=0.050$~s$^{-1}$)
lies below the plotted range.}
\label{fig:convergence}
\end{figure}

\section{Discussion}
\label{sec:disc}

\subsection{Positioning relative to existing literature}

Sakamoto \textit{et al.}~\cite{Sakamoto2024} demonstrated LGI violation
in closed-system LMG dynamics in principle.
The present paper extends this to the open-system regime and establishes
strict quantitative conditions: the violation is robust at $\gamma_\phi=0.05$~s$^{-1}$
($K_3\approx1.32$, $T_2^{\rm phys}=7.04$~ms) and requires
$\gamma_\phi\lesssim 0.289$~s$^{-1}$ ($T_2^{\rm phys}\gtrsim 1.22$~ms).
While exact closed-system dynamics inherently include multi-level effects,
standard open-system treatments typically rely on a mean-field two-level
approximation.
Relative to this standard mean-field baseline, the present paper makes five key additions.
First, exact parity protection yields the $2.35\times$ threshold enhancement
(Level~A to Level~B, Remark~\ref{rem:hierarchy}): because $\langle E_i|\hat{J}_z|E_i\rangle=0$
by the emergent collective $\mathbb{Z}_2$ symmetry, the spurious cross-term
that inflates the mean-field dephasing rate is eliminated exactly.
Second, the net contribution of higher odd-parity Dicke states yields
a further $2.47\times$ enhancement (Level~B to Level~C): the same selection
rule protects the coherences of $|E_3\rangle$, $|E_5\rangle$,\ldots,
and at the benchmark parameters their dynamical phases are constructive,
strengthening the $K_3$ correlator.
Third, both local and collective bath decoherence models are evaluated explicitly,
establishing the collective bath as the relevant model for BEC global
field noise.
Fourth, the Landau--Zener crossover provides a complementary, experimentally
direct quantum-classical discriminator.
Fifth, complete self-tested executable code is provided, reproducing all figures
and thresholds.

\subsection{The Goldilocks double squeeze}

The Goldilocks zone is intrinsically narrow because the system is squeezed
from both sides simultaneously.
For $N\ll N_c$: $\DeltaE\gg k_BT$, the system is deep in the quantum
regime but $\chi_{\rm eff}\sim N$ is small.
For $N\gg N_c$: thermal fluctuations overwhelm coherence.
For $N\gg N_c$ and collective dephasing: $T_2^{\rm coll}$ drops as
$1/N^2$, collapsing the $\Pket$--$\Rket$ superposition quadratically
faster than any linear gain in susceptibility.
The net result is a favourable operating window at $N\approx250$--$300$
under the benchmark BEC parameters used here; this window is genuinely
mesoscopic (neither microscopic nor macroscopic), though its precise
boundaries depend on the experimental noise floor and should be
calibrated via exact diagonalisation for any specific apparatus.
This range is defined by the condition $k_BT \lesssim \DeltaE(N) \lesssim 3k_BT$
(from the Goldilocks N-scan in Appendix~\ref{app:code}: at $N=250$, $\DeltaE\approx3338$~rad/s $\approx2.5k_BT$; at $N=300$, $\DeltaE\approx2299$~rad/s $\approx1.75k_BT$; at $N=370$, $\DeltaE=k_BT$ exactly).
The lower bound ($N\approx250$) ensures sufficient macroscopic susceptibility $\chi_{\rm eff}\sim N$; the upper bound ($N\lesssim350$) keeps $T_2^{\rm coll}>T_2^{\rm phys}$ and $K_3>1$ with robust margin.

\subsection{Closing the clumsiness loophole}

A quantum non-demolition (QND) measurement of $\hat{J}_z$ via Faraday rotation phase-contrast
imaging commutes with $\hat{J}_z$ up to quantum back-action and adds
no energy.
Non-invasiveness can be addressed by repeating measurements on the same
condensate and checking consistency.
Current technology achieves single-shot sensitivity $\Delta J_z\sim0.5$
with $10\,\mu$s integration~\cite{Muessel2014}, far shorter than
$T_2^{\rm phys}=1.22$~ms at the violation threshold.
The finite resolution $\Delta J_z\approx0.5$ does not compromise the
step-function observable $Q=\operatorname{sgn}(\hat{J}_z)$: the two
macroscopic phases $\Pket$ and $\Rket$ are separated by $Nm_*/2\approx57$
spins, giving a signal-to-noise ratio $Nm_*/(2\Delta J_z)\approx114\gg1$.
The measurement uncertainty overlaps the $J_z=0$ boundary by
$\sim\operatorname{erfc}(57/0.5)$, which is negligible in any physical apparatus.
At typical far-detuned probe parameters, the spontaneous Raman
scattering rate per atom is $\lesssim10^2$~s$^{-1}$~\cite{Muessel2014},
giving $\lesssim10^{-3}$ scattering events per atom during the
$10\,\mu$s integration ($121\times$ shorter than $T_2^{\rm phys}$),
confirming that the measurement is QND on the timescale of the
required coherence;
the loophole may be addressable with current QND techniques, though a detailed experimental error analysis remains necessary.

\subsection{Experimental protocol}

\textbf{For P4:} prepare a spinor BEC at $\Gamma/J=0.95$,
$N\approx250$--$300$, with a large initial longitudinal bias $h_{\rm init} \ll 0$ so the system is fully localised in the initial ground-state well $\Rket$. 
Apply a linear $h$-quench (laser detuning sweep) at rate $v_h$ through the degeneracy point $h=0$ to a final bias $h_{\rm final} \gg 0$.
Vary $\tau_Q$ from $10^{-4}\,\text{s}$ to $\sim10 T_2^{\rm phys}$ and measure
$P_{\rm error}$ by spin-resolved absorption imaging~\cite{Zibold2010}.
The classical prediction is $P_{\rm error}\approx1$ throughout
(since $\tau_K\gg\tau_Q$); the quantum prediction is exponential
decay to zero for $\tau_Q\gtrsim\hbar/\DeltaE$, providing a clear
binary discrimination without requiring knowledge of $T$.

\textbf{For P5:} prepare the same BEC in the symmetric ground state
$\alpha\Pket+\beta\Rket$.
Perform three sequential QND (quantum non-demolition) measurements at times $t_1<t_2<t_3$
with equal separation $\Delta t=\pi\hbar/(3\DeltaE)\approx0.8$~ms.
Compute $K_3$ over an ensemble.
Target: $\gamma_\phi\lesssim 0.289$~s$^{-1}$, which current BEC experiments
comfortably achieve.

\subsection{Limitations}
\label{sec:limitations}

The two-level LZ and LGI formulae are qualitative near criticality
(Remark~\ref{rem:two_level}).
The instanton prefactor $C_0$ introduces $\mathcal{O}(1)$ uncertainty
in $N_c$; exact diagonalisation is recommended for experimental planning.
A full parameter sweep over $N$ and $\Gamma/J$ using the code in
Appendix~\ref{app:code} is the appropriate next step.

The Lindblad master equation explicitly breaks the unitarity of the
system alone, which is its physical content.
The robustness of $K_3>1$ against any proposed non-unitary corrections
to quantum mechanics (GRW, CSL, Penrose collapse) is quantified in the
Conclusion (Section~\ref{sec:conclusion}), where it is identified as a
fourth structural feature of the protocol: the corrections would add
effective dephasing $\lesssim N_{\rm nuc}\lambda_{\rm GRW}\approx3\times10^{-12}$~s$^{-1}$
(using $N_{\rm nuc}=370\times87=32{,}190$ nucleons for $^{87}$Rb at
$\lambda_{\rm GRW}\sim10^{-16}$~s$^{-1}$ per nucleon~\cite{Emary2014}),
more than $10^{10}$ times below the experimental $\gamma_\phi=0.05$~s$^{-1}$.
One genuine experimental caveat beyond the Lindblad model is atom-number
fluctuations.
The instanton formula $\DeltaE\propto N^{1/2}e^{-NS_{\rm inst}}$ gives
the logarithmic derivative
$d(\ln\DeltaE)/dN = 1/(2N)-S_{\rm inst}$,
which at $N=370$, $S_{\rm inst}=0.010787$ evaluates to
$1/740-0.010787=-0.00944$~atom$^{-1}$.
For shot-noise fluctuations $\delta N\sim\sqrt{N}\approx19.2$ atoms:
\begin{equation}
  \frac{\delta(\DeltaE)}{\DeltaE}
    = \left|\frac{d\ln\DeltaE}{dN}\right|\delta N
    = 0.00944\times19.2 \approx 18\%.
\end{equation}
This is confirmed by exact diagonalisation:
$d\DeltaE/dN|_{N=370}\approx-10.95$~rad/s per atom, giving
$\delta(\DeltaE)=10.95\times19.2\approx210$~rad/s $\approx16\%$
(the $\sim$10\% difference from the instanton estimate reflects
the $\mathcal{O}(1)$ prefactor uncertainty).
These fluctuations inhomogeneously
average $K_3$ over the ensemble and should be accounted for in a full
experimental analysis.

Finally, the collective $\mathbb{Z}_2$ symmetry protecting the LGI correlator from $T_1$ mixing is exact within the LMG model, but the two-mode BEC approximation breaks down at $\mathcal{O}(1/N)$ due to scattering into higher momentum modes. Full beyond-mean-field simulations or experimental calibration are the natural next steps.
For $^{87}$Rb at typical BEC densities, the three-body recombination
lifetime is $\mathcal{O}(\text{seconds})$~\cite{Trenkwalder2016};
over the $\sim2.5$~ms total LGI protocol, expected atom loss is
$\ll1$ atom, so $N$ is effectively conserved and the tunnel splitting
$\DeltaE$ does not drift during the measurement sequence.

\begin{remark}[Measurement back-action and five-level truncation]
The five-level Lindblad simulation truncates to the lowest five Dicke
states and projects $Q=\operatorname{sgn}(\hat{J}_z)$ into this subspace.
In a real BEC, a projective measurement of $Q$ produces back-action
that in principle scatters population into higher Dicke levels outside
the truncated manifold.
This heating is suppressed here because $Q_{01}^2=0.9436$ is close to
unity: the overwhelming majority of probability amplitude ($>94\%$)
remains in the lowest two levels after each measurement, so the
higher-level leakage is a small correction.
Moreover, as shown above ($Q_{00}=Q_{11}=0$), the $K_3$ correlator is
immune to $T_1$ population mixing between $|E_0\rangle$ and $|E_1\rangle$,
since $Q$ acts only off-diagonally in the parity basis.
This immunity is exact within the idealised LMG parity symmetry at $h=0$;
beyond-two-mode corrections at $\mathcal{O}(1/N)$ from scattering into higher
momentum modes may introduce $\sim$percent-level sensitivity in a physical BEC,
and should be assessed in a full beyond-mean-field simulation.
\end{remark}
\section{Conclusion}
\label{sec:conclusion}

We have identified one strictly model-independent classically forbidden
signature (P5) and one model-dependent quantum--classical discriminator (P4) of mesoscopic LMG dynamics: the Landau--Zener adiabatic error rate exhibiting quantum
tunnelling ($P_{\rm error}\to0$) versus classical non-ergodic trapping
($P_{\rm error}\to1$) (P4), and the Leggett--Garg correlator $K_3>1$ (P5).
The quantum--classical separation in P4 is parametrically large and does
not rely on a thermal equilibrium comparison; it exploits the fact that
macroscopic classical escape ($\tau_K$) is exponentially slower than the quench
timescale ($\tau_Q\sim\text{ms}$), so the classical system is kinetically
frozen while the quantum system tunnels coherently.

The LGI threshold exists at three successive approximation levels
(Remarks~\ref{rem:hierarchy} and \ref{rem:reconcile}).
The mean-field two-level formula (level A, Key Result box) gives
$T_2^{\rm coll}\gtrsim3.93\hbar/\DeltaE$, corresponding to
$\gamma_\phi^{\rm (A)}\lesssim0.050$~s$^{-1}$; at the BEC target
$\gamma_\phi=0.05$~s$^{-1}$ this level predicts $K_3\approx1.000$
(the system sits exactly at the mean-field threshold).
The analytic eigenstate formula (level B), which correctly drops the
parity-forbidden cross-term $\gamma_\phi(Nm_*/2)^2\approx3337\gamma_\phi$,
raises the threshold $2.35\times$ to $\gamma_\phi^{\rm (B)}\lesssim0.117$~s$^{-1}$,
with $K_3\approx1.22$ at the BEC target and decoherence-free ceiling $K_3^{\rm max}=1.415$.
The five-level Lindblad simulation (level C, $Q_{01}^2=0.9436$) raises it
a further $2.47\times$ to $\gamma_\phi^{\rm (C)}\lesssim0.289$~s$^{-1}$
via reinforcing contributions from higher odd-parity Dicke states,
with $K_3\approx1.32$ at the BEC target; the two-level decoherence-free
ceiling $K_3^{\rm max}=Q_{01}^2\times\tfrac{3}{2}=1.415$ is shared by all
two-level descriptions at $T_2\to\infty$ and is not a level-C-specific prediction.
The overall A$\to$C improvement is $5.8\times$; level~(C) is the experimental threshold for BEC planning.
At the BEC target $\gamma_\phi=0.05$~s$^{-1}$, the five-level calculation
gives $K_3\approx1.32$ and $T_2^{\rm phys}\approx 7.04$~ms --- a robust,
quantitatively substantial violation with an $8.8\times$ margin above
the optimal measurement interval.
The decoherence-only size bound is $N\lesssim 890$, but the physically
operative constraint is the Goldilocks condition, restricting the
optimal range to $N\approx250$--$300$.

All results are testable in current spinor BEC experiments.
The Python code in Appendix~\ref{app:code} is self-tested and reproduces
all data points in Fig.~\ref{fig:multilevel} and the threshold
Eq.~\eqref{eq:gamma_thresh}; the Goldilocks crossover values in
Table~\ref{tab:goldilocks} require separate exact diagonalisation
over a range of $N$ as described in Section~\ref{sec:goldilocks}.

Four structural features of the LGI result deserve emphasis, as they
bear directly on the physical interpretation of the protocol and on
its forthcoming extension to a broader theoretical framework.

First, the entire quantitative structure of the LGI protocol rests on the interplay between an \emph{emergent intrinsic symmetry} and a \emph{tailored operational measurement choice}.

\smallskip
\noindent\textbf{Intrinsic symmetry (system and environment).}
The collective $\mathbb{Z}_2$ parity generated by $\hat{P}=e^{i\pi\hat{J}_x}$
is respected by both the LMG Hamiltonian at $h=0$ and the collective bath
$L_z=\sqrt{\gamma_\phi}\hat{J}_z$.
(i)~It dictates $\langle E_i|\hat{J}_z|E_i\rangle=0$ exactly for all $i$,
eliminating the spurious cross-term that inflates the mean-field dephasing
rate and yielding the $2.35\times$ Level~A$\to$B improvement.
(ii)~The same selection rule protects the coherences of higher odd-parity
states ($\langle E_3|\hat{J}_z|E_3\rangle=0$); whether each contributes
constructively or destructively to $K_3$ depends on the dynamical phase
$\cos(\Delta E_{k0}\Delta t^*/\hbar)$. At the benchmark parameters the
phases of $|E_3\rangle$ and $|E_5\rangle$ are constructive, producing
the $2.47\times$ Level~B$\to$C improvement.
This last point is a numerical property of the spectrum at these
parameters, not a structural consequence of the symmetry.

\smallskip
\noindent\textbf{Measurement choice (protocol design).}
By choosing $Q=\operatorname{sgn}(\hat{J}_z)$, which is strictly odd
under $\hat{P}$, the protocol exploits this symmetry directly.
(iii)~$Q_{00}=Q_{11}=0$ exactly, making the LGI correlator blind to
$T_1$-type population mixing and requiring no ground-state preparation.
(iv)~Measurement back-action scatters no amplitude from $|E_0\rangle$
into even-parity states ($Q_{20}=Q_{40}=0$ exactly), confining spectral
leakage to odd states ($|E_5\rangle$: $0.45\%$, $|E_7\rangle$: $0.26\%$,
rapidly decreasing), which validates the five-level truncation.
All four consequences collapse simultaneously under local spin noise
$L_z^{(i)}=\sqrt{\gamma_\phi}\hat\sigma_i^z$, which breaks parity.

\smallskip
\noindent
Finally, the stationarity assumption $C_{23}=C_{12}$ rests on one
additional operational property independent of symmetry:
near-dichotomicity $Q^2\approx I$, holding up to $0.24\%$ corrections
from the $m=0$ Dicke state (absorbed into $Q_{01}^2=0.9436<1$).

Second, the theoretical prediction uses the standard operational quantum
description of projective measurements: the Lindblad master equation,
the Born rule for outcome probabilities, and the Lüders non-selective
instrument $\frac{1}{2}\{Q,\rho(t_i)\}$, formulated at the density-matrix
level without invoking selective collapse in individual runs.
Decoherence --- the continuous, non-selective entanglement of the system
with its environment, governed by $\gamma_\phi$ --- is explicitly
included and is the quantity being experimentally constrained.
Selective wavefunction collapse (instantaneous and interpretation-dependent)
is not assumed; the density matrix formalism is sufficient throughout.

Third, the violation $K_3>1$ is interpretation-independent: it is a
directly observable algebraic inequality on measured temporal
correlators, and its violation rules out all macrorealist models
simultaneously regardless of which interpretation of quantum mechanics
one adopts.\footnote{%
Interpretational nuances arise at two points.
(i)~\emph{Non-invasive measurability (NIM):}
the classical LGI bound requires that measurements do not disturb a
system that is in a definite classical state.
In the BEC context this is addressed by the QND protocol
(Section~\ref{sec:disc}), which commutes with $\hat{J}_z$ and adds no
energy; in Copenhagen the measurement does disturb the quantum system,
but this disturbance is precisely the violation mechanism, not a loophole.
(ii)~\emph{Meaning of the violation:}
in Copenhagen, $K_3>1$ signals the absence of pre-existing definite
values of $Q$ before measurement; in many-worlds, it reflects
quantum entanglement between branches; in QBism, it updates an agent's
belief state.
All interpretations agree on the predicted and observed value of $K_3$;
they disagree only on the ontological description of what occurs between
measurements.
The Lindblad equation and Born rule are shared by all interpretations
as computational tools, so the theoretical predictions of this paper
are fully interpretation-neutral.}

Fourth, the violation is robust against any known proposal for
non-unitary corrections to quantum mechanics beyond the Lindblad model.
The Lindblad dissipator explicitly breaks the unitarity of the system
alone --- this is its physical content, not a limitation --- and the
prediction $K_3>1$ requires only Hilbert space structure, the Born
rule, and CPTP dynamics; it does not require exact global unitarity.
Spontaneous-collapse models (GRW, CSL) and Penrose gravitational
collapse all predict additional effective dephasing bounded by
$\lesssim N_{\rm nuc}\lambda_{\rm GRW}\approx3\times10^{-12}$~s$^{-1}$
(for $N=370$ atoms of $^{87}$Rb, using $\lambda_{\rm GRW}\sim10^{-16}$~s$^{-1}$
per nucleon and $N_{\rm nuc}=370\times87=32{,}190$),
more than $10^{10}$ times below the dominant experimental dephasing
$\gamma_\phi=0.05$~s$^{-1}$.
The violation is therefore protected not only against the environmental
noise we have quantified, but against the strongest speculative
modifications to quantum mechanics currently proposed: the result is
insensitive to whether quantum mechanics is exactly unitary at the
fundamental level, as long as any additional non-unitary corrections
remain within the bounds set by existing collapse models.

These four features --- emergent symmetry protection, collapse-free
measurement theory, interpretation independence, and robustness against
beyond-quantum corrections --- together make the LGI violation in the
LMG model structurally well-protected against the dominant experimental
decoherence channels, providing a quantitatively grounded and
experimentally actionable test of macrorealism.
All four are established in this paper; a forthcoming companion paper
applies the same structural framework in a broader context.

\subsection*{Declaration of AI-Assisted Technologies}
During the preparation of this work, the author used Claude (Anthropic) and Gemini Studio
for the purpose of coding numerical simulations, generating figures,
and manuscript preparation assistance.
After using these tools, the author reviewed and edited the output as needed
and takes full responsibility for the content and scientific accuracy
of the published article.

\appendix

\section{Mean-Field Commutator Derivation}
\label{app:mf}

The mean-field Hamiltonian per spin is
$\hat{H}_{\rm MF}=-(Jm_z+h)\hat\sigma^z-\Gamma\hat\sigma^x$.
Using $[\hat\sigma^x,\hat\sigma^z]=-2i\hat\sigma^y$:
\begin{align}
  \frac{d\langle\hat\sigma^z\rangle}{dt}
    &= i\langle[-\Gamma\hat\sigma^x,\hat\sigma^z]\rangle
     = -2\Gamma\langle\hat\sigma^y\rangle,\\
  \frac{d\langle\hat\sigma^y\rangle}{dt}
    &= -2(Jm_z+h)\langle\hat\sigma^x\rangle+2\Gamma\langle\hat\sigma^z\rangle,\\
  \frac{d\langle\hat\sigma^x\rangle}{dt}
    &= 2(Jm_z+h)\langle\hat\sigma^y\rangle.
\end{align}
Under the mean-field approximation all spins are equivalent,
so $\langle\hat\sigma_i^\alpha\rangle=\langle\hat\sigma^\alpha\rangle$
for every $i$ and every component $\alpha\in\{x,y,z\}$.
Summing over all $N$ spins and applying the definition
$m_\alpha\equiv N^{-1}\sum_i\langle\hat\sigma_i^\alpha\rangle$
then gives $m_\alpha=\langle\hat\sigma^\alpha\rangle$ directly
(the factor of $N$ from the sum and the $1/N$ in the definition cancel),
reproducing Eq.~\eqref{eq:bloch}.

\section{Linearised ACF: Exact Normal-Mode Frequency and Damping}
\label{app:acf}

The stationary autocorrelation function of the longitudinal magnetisation
is defined as
\begin{equation}
  C_{zz}(\Delta t) = \langle\delta m_z(t+\Delta t)\,\delta m_z(t)\rangle_{\rm st},
  \qquad \delta m_z \equiv m_z - m_z^{\rm eq}.
\end{equation}

\textbf{True fixed point.}

The true deterministic fixed point in the ordered phase ($\Gamma<J$, $h=0$)
is the tilted vector
\begin{equation}
  m_x^{\rm eq} = \frac{\Gamma}{J}, \qquad
  m_y^{\rm eq} = 0, \qquad
  m_z^{\rm eq} = m_* = \sqrt{1-\frac{\Gamma^2}{J^2}},
  \label{eq:fp_ordered}
\end{equation}
which satisfies $|\mathbf{m}^{\rm eq}|=1$ and all three Bloch equations
identically~\cite{Risken1989}.

\textbf{Linearisation around the correct fixed point.}
Setting $\delta m_\alpha = m_\alpha - m_\alpha^{\rm eq}$ and expanding
to first order, the cross terms coupling $\delta m_z$ to the transverse
subsystem cancel exactly in the $\dot{m}_y$ equation
(since $-2J(m_x^{\rm eq}) + 2\Gamma = -2\Gamma + 2\Gamma = 0$).
Assuming phenomenological relaxation directed towards the instantaneous local equilibrium (i.e., damping proportional to $-(m_\alpha - m_\alpha^{\mathrm{eq}})$, which preserves the true tilted fixed point unlike standard $T_2$ decay to zero), this leaves the closed $(\delta m_x, \delta m_y)$ subsystem:
\begin{align}
  \frac{d}{dt}\delta m_x &= +2Jm_*\,\delta m_y - \Gamma_2\,\delta m_x,\\
  \frac{d}{dt}\delta m_y &= -2Jm_*\,\delta m_x - \Gamma_2\,\delta m_y,\\
  \frac{d}{dt}\delta m_z &= -2\Gamma\,\delta m_y - \Gamma_1\,\delta m_z.
  \label{eq:lin_bloch}
\end{align}
The linearised matrix has eigenvalues
$\lambda_{1,2}=-\Gamma_2\pm i\,\omega_0$ and $\lambda_3=-\Gamma_1$,
with
\begin{align}
  \omega_0 &= 2Jm_*, \label{eq:omega0}\\
  \gamma_{\rm osc} &= \Gamma_2 = 1/T_2^{\rm cl}. \label{eq:gamma_acf}
\end{align}
The oscillatory modes decay at \emph{exactly} $\Gamma_2$; the longitudinal
mode decays at $\Gamma_1$.
Applying the regression theorem (Risken~\cite{Risken1989}, Sect.~3.6)
to this two-component structure gives:
\begin{equation}
  \frac{C_{zz}(\Delta t)}{C_{zz}(0)}
    = A_0\,e^{-\Gamma_2\Delta t}\cos(\omega_0\Delta t)
      + B_0\,e^{-\Gamma_1\Delta t},
  \label{eq:acf}
\end{equation}
where $A_0$ and $B_0$ depend on the stationary variance and the
geometric projection of $\delta m_z$ onto the two modes.
For $\Delta t\ll T_1$ (since $T_1\gg T_2^{\rm cl}$ in spinor BECs), the slow
$B_0 e^{-\Gamma_1\Delta t}$ baseline is negligible and the ACF is
dominated by the oscillatory term.
Numerically verified at $J=1$, $\Gamma=0.5$, $\Gamma_2=0.1$,
$\Gamma_1=0.01$: the linearised matrix eigenvalues are
$-0.1\pm1.732i$ and $-0.01$, confirming $\omega_0=2Jm_*=\sqrt{3}\approx1.732$
and $\gamma_{\rm osc}=\Gamma_2=0.1$ to five significant figures.
In the deep ordered phase ($m_*\to1$): $\omega_0\to2J$;
near criticality ($m_*\to0$): $\omega_0\to0$ (critical slowing down).

\section{Freeze-out Scaling: KZM (J-quench) and LZ Bias-Sweep (h-quench)}
\label{app:kzm}

The Kibble--Zurek freeze-out time scales as
$\hat{t}\propto\tau_Q^{\mu/(1+\mu)}$~\cite{Zurek1996,Defenu2018,Puebla2020},
where $\mu$ is the dynamical critical exponent defined by
$\tau_{\rm rel}\propto|\epsilon|^{-\mu}$ and $\epsilon$ is the
dimensionless distance from the critical point.
The two physically distinct quench protocols are treated separately:
the J-quench passes through a genuine quantum critical point and
falls within the standard KZM framework; the h-quench operates within
the already-ordered phase and is physically an avoided level crossing
(LZ bias-sweep), not a critical phenomenon.

\textbf{J-quench} ($\epsilon_J\equiv J/\Gamma-1$, quenching from
disordered to ordered phase through $\Gamma_c=J$).
Linearising the undamped Bloch equations~\eqref{eq:bloch} around the
disordered fixed point $(m_x,m_y,m_z)=(1,0,0)$ at $h=0$, the
transverse perturbations $(\delta m_y,\delta m_z)$ obey:
\begin{align}
  \frac{d}{dt}\delta m_y &= \phantom{-}2(\Gamma-J)\,\delta m_z, \nonumber\\
  \frac{d}{dt}\delta m_z &= -2\Gamma\,\delta m_y. \nonumber
\end{align}
Combining gives $\frac{d^2}{dt^2}\delta m_z = -\omega_J^2\,\delta m_z$ with
$\omega_J^2 = 4\Gamma(\Gamma-J) = 4\Gamma^2|\epsilon_J|$, hence
$\omega_J\propto|\epsilon_J|^{1/2}$ and $\mu_J=1/2$, giving:
\begin{equation}
  \hat{t}_{\rm J}\propto\tau_Q^{(1/2)/(3/2)} = \tau_Q^{1/3}.
  \label{eq:kzm_j}
\end{equation}

\textbf{h-quench: LZ bias-sweep scaling} ($\epsilon_h\equiv h$,
sweeping the longitudinal bias in the already-ordered phase $\Gamma<J$).
This is \emph{not} a standard KZM transition: the h-quench does not
pass through a quantum critical point, and for finite $N$ it is an
avoided level crossing with tunnel splitting $\Delta E$ (LZ physics);
in the thermodynamic limit it becomes a first-order transition with
no diverging correlation length.
The $\tau_Q^{1/2}$ scaling derived below is therefore LZ bias-sweep
scaling, not KZM universality.
In the ordered phase at $h=0$, the two vacua $\Pket$ and $\Rket$
are degenerate with tunnel splitting $\DeltaE$.
A nonzero $h$ couples linearly to the order parameter $m_z$,
splitting the vacuum energies as $E_\pm\approx\mp Nhm_*$;
the gap therefore closes \emph{linearly}:
\begin{equation}
  \Delta(h)\approx 2Nhm_* + \DeltaE \approx 2Nhm_*
  \quad(h\gg \DeltaE/2Nm_*),
  \label{eq:gap_h}
\end{equation}

giving $\omega_h\propto|\epsilon_h|^1$ and $\mu_h=1$:
\begin{equation}
  \hat{t}_{\rm h}\propto\tau_Q^{1/(1+1)} = \tau_Q^{1/2}.
  \label{eq:kzm_h}
\end{equation}
This is \emph{identical} to the classical overdamped exponent
($\mu_{\rm class}=1$, $\hat{t}\propto\tau_Q^{1/2}$), which follows
from $\tau_{\rm rel}^{\rm class}\propto|\epsilon|^{-1}$ for a
bistable potential near its bifurcation.
The h-quench is therefore non-discriminating between quantum and
classical dynamics on the basis of the KZM exponent alone;
P4 and P5 are required to break this degeneracy.

\begin{table}[H]
\centering
\small
\caption{Freeze-out scaling exponents.
The J-quench is standard KZM (critical point, diverging correlation length).
The h-quench is LZ bias-sweep scaling (avoided level crossing in ordered
phase, no critical point): the $\tau_Q^{1/2}$ exponent matches the
classical overdamped case for distinct physical reasons.}
\label{tab:kzm}
\setlength{\tabcolsep}{8pt}
\begin{tabular}{@{}lccc@{}}
\toprule
Scenario & $\omega\propto|\epsilon|^?$ & $\mu$ & $\hat{t}\propto\tau_Q^?$\\
\midrule
Classical overdamped & --- & $1$ & $1/2$\\
Quantum J-quench     & $|\epsilon|^{1/2}$ & $1/2$ & $1/3$\\
Quantum h-quench     & $|\epsilon|^{1}$   & $1$   & $1/2$\\
\bottomrule
\end{tabular}
\end{table}

\section{LGI Classical Bound}
\label{app:lgi}

Under macrorealism (MR) plus non-invasive measurability (NIM), each
trial assigns definite values $Q_i\in\{+1,-1\}$ to the observable
at each time, regardless of measurement~\cite{LeggettGarg1985,Emary2014}.
Exhaustive enumeration of $2^3=8$ sign combinations shows
$Q_1Q_2+Q_2Q_3-Q_1Q_3\leq1$ in every case; taking the expectation
value over any classical probability distribution: $K_3\leq1$.
This bound holds for \emph{all} macrorealist models satisfying MR+NIM,
including non-Markovian classical stochastic processes with arbitrary
memory, provided the NIM assumption is not violated.
The bound is violated by any quantum system whose off-diagonal coherences
between the $Q=\pm1$ subspaces survive to the measurement time.

The threshold Eq.~\eqref{eq:t2req} is evaluated at
$\Delta t=\pi\hbar/(3\DeltaE)$, which is the exact optimum only at
$T_2=\infty$.
For finite $T_2$ the true optimal $\Delta t$ is slightly smaller,
making Eq.~\eqref{eq:t2req} conservative.

\section{Thermodynamically Consistent Fokker--Planck Dynamics}
\label{app:fp}

Model~A is used here as a thermodynamically consistent phenomenological
description of classical barrier crossing, not as a microscopic
derivation from the LMG master equation or from the Lindblad dynamics.
Specifically, it is a \emph{separate classical foil} introduced to
contrast with the quantum LMG: the classical Kramers activation is
governed by the $T=10$~nK motional bath of the BEC gas (obeying the
fluctuation-dissipation theorem), while the quantum dephasing bounded
in the main text is driven by magnetic-field noise acting as a
pure-dephasing environment; these are physically distinct environments
(see Section~\ref{sec:lindblad}).

To correctly capture thermal escape across the macroscopic barrier, the
dynamics must respect the thermodynamic free energy landscape
$U(m_z)=Nf(m_z)$, where $f(m_z)$ is the intensive mean-field free
energy.
Rather than relying on phenomenological $T_1/T_2$ Bloch equations ---
which do not strictly obey detailed balance with respect to the
interacting nonlinear free energy~\cite{Hohenberg1977} --- we adopt a
thermodynamically consistent stochastic approach.
We model the macroscopic variable $m_z$ via an overdamped Langevin
equation (Model~A dynamics) governed by the free energy gradient:
\begin{equation}
  \frac{d m_z}{dt} = -\Gamma_{\rm eff} \frac{\partial f(m_z)}{\partial m_z} + \xi(t),
  \label{eq:langevin}
\end{equation}
where $\Gamma_{\rm eff}$ is an effective phenomenological mobility
treated as a free parameter.
A heuristic quantum-Zeno scaling argument (not a rigorous derivation
from the LMG master equation) suggests
$\Gamma_{\rm eff}\propto4\Gamma^2/\Gamma_2\propto T_2^{\rm cl}$,
implying stronger transverse decoherence suppresses longitudinal
mobility; this motivates the Model~A foil, but the actual value of
$\Gamma_{\rm eff}$ is not used in any quantitative prediction.

The stochastic forcing $\xi(t)$ is zero-mean Gaussian white noise
representing coupling to the $T=10$~nK motional bath:
\begin{equation}
  \langle\xi(t)\rangle = 0, \qquad
  \langle\xi(t)\xi(t')\rangle = 2D\,\delta(t-t').
\end{equation}
To ensure relaxation to the correct thermal equilibrium distribution
$P_{\rm eq}\propto\exp(-Nf(m_z)/k_BT)$, the diffusion coefficient $D$
and mobility $\Gamma_{\rm eff}$ must satisfy the Einstein relation
(fluctuation-dissipation theorem):
\begin{equation}
  D = \frac{\Gamma_{\rm eff} k_B T}{N}.
  \label{eq:fdt}
\end{equation}
The $1/N$ scaling of $D$ reflects that $m_z=M_z/N$ is an intensive
variable subject to macroscopic thermodynamic constraints.

The equivalent Fokker--Planck equation follows directly from standard
stochastic calculus~\cite{Risken1989}:
\begin{equation}
  \frac{\partial P}{\partial t} = \frac{\partial}{\partial m_z}\!\left[
    \Gamma_{\rm eff} \frac{\partial f(m_z)}{\partial m_z} P
  \right] + D \frac{\partial^2 P}{\partial m_z^2}.
  \label{eq:fp_derived}
\end{equation}
This separates the physical forces cleanly: the deterministic drift
$-\Gamma_{\rm eff}f'(m_z)$ drives the system toward the ordered minimum
$m_z=m_*$, while the diffusion term $D\,\partial^2P/\partial m_z^2$
represents thermal fluctuations that occasionally drive the system over
the barrier.
Because $D\propto1/N$, the restoring force dominates in the
thermodynamic limit.
The Kramers time scales as
$\tau_K\propto\Gamma_{\rm eff}^{-1}e^{N\Delta f_0/k_BT}$; at
$\Gamma/J=0.95$, $N=370$, $T=10$~nK:
$N\Delta f_0/k_BT\approx13.1$, giving
$e^{N\Delta f_0/k_BT}\approx5\times10^5$, and with attempt period
$2\pi/\omega_0\approx0.27$~ms one obtains $\tau_K\sim\mathcal{O}(10^2)$~s,
vastly exceeding any accessible sweep timescale $\tau_Q\sim\text{ms}$.
\section{Mean First-Passage Time (MFPT) via Kramers Escape Rate}
\label{app:mfpt}

With the extensive effective potential $U(m_z)=Nf(m_z)$ and the intensive diffusion coefficient $D=\Gamma_{\rm eff}k_BT/N$, the classical Kramers escape rate~\cite{Kramers1940} across the symmetric barrier yields the mean first-passage time.
Applying the standard result $\tau_K = 2\pi k_BT / (D\cdot N\sqrt{f''(m_*)|f''(0)|}) \cdot e^{N\Delta f_0/k_BT}$ and substituting $D = \Gamma_{\rm eff}k_BT/N$:
\begin{equation}
  \langle\tau\rangle\xrightarrow{h\to0}
    \tau_K = \frac{2\pi}
    {\Gamma_{\rm eff}\sqrt{f''(m_*)|f''(0)|}}
    \,e^{N\Delta f_0/k_BT}.
  \label{eq:kramers}
\end{equation}
Notably, the macroscopic system size $N$ cancels exactly in the prefactor---the $\mathcal{O}(N)$ scaling of the potential curvature perfectly offsets the $\mathcal{O}(1/N)$ scaling of the intensive mobility---confining the $N$-dependence strictly to the exponential.

Here, $\Delta f_0\equiv f(0)-f(m_*)>0$ is the mean-field free-energy barrier per spin,
evaluated from Eq.~\eqref{eq:freeenergy} at $h=0$.
Using the large-argument approximation $\ln\cosh(x)\approx x-\ln 2$
(valid for $x\gg1$; here $J/T\approx28.4$ and $\Gamma/T\approx27.0$):
\begin{align}
  f(0) &\approx -\Gamma + T\ln 2, \nonumber\\
  f(m_*) &\approx \tfrac{J}{2}m_*^2 - J + T\ln 2, \nonumber\\
  \Delta f_0 &\approx -\tfrac{J}{2}m_*^2 + J - \Gamma. \nonumber
\end{align}
At $J=37{,}195$~rad/s, $\Gamma/J=0.95$, $m_*^2=0.0975$:
$\Delta f_0\approx 46.5$~rad/s, giving
$N\Delta f_0/k_BT = 370\times46.5/1310 = 13.1$.

In the ordered phase ($\Gamma<J$), the double-well structure of $f(m_z)$ guarantees $f(0)>f(m_*)$, ensuring $\Delta f_0>0$ and causing $\tau_K$ to diverge exponentially with $N$. Algebraic divergence of $\langle\tau\rangle$ as $h\to0$ occurs only at the critical point $\Gamma=J$, where $\Delta f_0\to0$ and the quadratic Kramers approximation gives way to critical slowing down.

The mean-field free energy $f(m_z)$ is exact only in the $N\to\infty$
limit; at $N=370$, $\mathcal{O}(1/N)$ corrections to the saddle-point
landscape modify $\Delta f_0$ by $\mathcal{O}(1/N)$ per spin, shifting
the exponent $N\Delta f_0/k_BT\approx13.1$ by $\mathcal{O}(1)$.
This changes $\tau_K$ by at most a factor of $e^{\pm1}\approx2$--$3$,
leaving it in the range $\mathcal{O}(10\text{--}10^3)$~s --- still five
orders of magnitude above $\tau_Q\sim1$~ms.
The parametric separation $\tau_K\gg\tau_Q$ that drives classical
non-ergodicity is robust against finite-$N$ corrections to the
mean-field landscape.

\section{WKB Instanton Action}
\label{app:instanton}

To analytically estimate the tunnel splitting $\Delta E(N) \propto e^{-N S_{\rm inst}}$ used to define the Goldilocks crossover in Section 5, we evaluate the WKB instanton action in the thermodynamic limit. 

Following standard semiclassical methods~\cite{Sachdev2011}, we map the LMG Hamiltonian to spin-coherent states $|\theta,\phi\rangle$ with $z=\cos\theta$. Analytically continuing to imaginary azimuthal momentum $\phi\to i\phi$, the classically forbidden region satisfies $\cosh\phi(z)=\frac{J(1-z^2)+\Gamma^2/J}{2\Gamma\sqrt{1-z^2}}$. Setting $x=\frac{J}{\Gamma}\sqrt{1-z^2}$ gives the momentum trajectory $\phi(z)=\ln x$, yielding the tunnelling integrand $\ln\!\bigl(\tfrac{J\sqrt{1-z^2}}{\Gamma}\bigr)$.

As detailed in standard quantum phase transition literature~\cite{Sachdev2011}, the leading contribution to the tunnel splitting arises from a single instanton trajectory connecting the two vacua ($-m_* \to +m_*$).
The WKB exponent for a spin-$J=N/2$ system involves $J\int_{-m_*}^{+m_*}[\ldots]dz$;
because the integrand $\ln\!\bigl(\tfrac{J\sqrt{1-z^2}}{\Gamma}\bigr)$ is \emph{even} in $z$, this equals
$\frac{N}{2}\times2\int_0^{m_*}[\ldots]dz = N\times S_{\rm inst}$.
The factor of $2$ from the barrier symmetry precisely cancels the $\tfrac{1}{2}$ from $J=N/2$,
which is why the intensive instanton action is defined by integrating from the barrier
top ($z=0$) to one vacuum ($z=m_*$):
\begin{equation}
  S_{\rm inst} = \int_0^{m_*}\ln\!\left(\frac{J\sqrt{1-z^2}}{\Gamma}\right)dz.
  \label{eq:S_half_integral}
\end{equation}
Using integration by parts ($u=\ln(\dots), dv=dz$), this integral evaluates exactly to $z \ln(\dots) - z + \operatorname{arctanh}(z)$. Applying the limits from $z=0$ to the vacuum condition $m_*=\sqrt{1-\Gamma^2/J^2}$ yields the closed-form expression:
\begin{equation}
  S_{\rm inst} = \operatorname{arctanh}(m_*)-m_*.
  \label{eq:S_half_closed}
\end{equation}

The full tunnel splitting takes the form $\DeltaE(N)=C_0N^{1/2}e^{-NS_{\rm inst}}$.
The $N^{1/2}$ prefactor originates from the time-translational zero-mode of the instanton~\cite{Coleman1985,Chudnovsky1998}: changing variables from the zero-mode fluctuation to the temporal centre introduces a Jacobian $\propto\sqrt{\mathcal{S}}\propto\sqrt{N}$. The coefficient $C_0\sim\hbar\omega_0$ is the attempt frequency from the Gaussian integral over remaining non-zero fluctuation modes.

The analytic roots extracted from this exponent ($S_{\rm inst}=0.000947$, $0.010787$, and $0.031255$ for $\Gamma/J=0.99$, $0.95$, and $0.90$ respectively) define the theoretical crossover boundaries in Table~\ref{tab:goldilocks} and are confirmed to match the exact diagonalisation condition $\Delta E/k_BT \approx 1$ at $N=370$.

\section{Decoherence Scaling Derivations}
\label{app:decoherence}

\textbf{Local bath.}
Approximating $\Pket\approx|{+}\rangle^{\otimes N}$ and
$\Rket\approx|{-}\rangle^{\otimes N}$, we evaluate the Lindblad
dissipator $\mathcal{D}[\rho]=\gamma_\phi\sum_i(\hat\sigma_i^z\rho\hat\sigma_i^z-\rho)$
on the off-diagonal coherence
$\rho_{PR}\equiv\langle P|\rho|R\rangle$.
For a single spin $i$, using $\langle+|\hat\sigma^z=\langle+|$
and $\hat\sigma^z|-\rangle=-|-\rangle$:
\begin{equation}
  \langle P|\hat\sigma_i^z\rho\hat\sigma_i^z|R\rangle
  = \langle+|\hat\sigma^z|+\rangle\,\rho_{PR}\,\langle-|\hat\sigma^z|-\rangle
  = (+1)\rho_{PR}(-1) = -\rho_{PR},
\end{equation}
while the two sandwich terms give
\begin{equation}
  -\tfrac{1}{2}\langle P|(\hat\sigma_i^{z\,2}\rho
    + \rho\hat\sigma_i^{z\,2})|R\rangle
  = -\tfrac{1}{2}(1+1)\rho_{PR} = -\rho_{PR},
\end{equation}
where $(\hat\sigma^z)^2=\hat{I}$ has been used.
Summing the two contributions per spin: $(-1-1)\rho_{PR}=-2\rho_{PR}$.

However, spin-coherent states at finite $m_*$ are \emph{not} Dicke
extremes: a spin-coherent state $|{\bf n}\rangle$ at polar angle $\theta$
(with $m_*=\cos\theta$) obeys
$\langle{\bf n}|\hat\sigma^z|{\bf n}\rangle=m_*=\cos\theta$, giving
\begin{equation}
  \langle P|\hat\sigma_i^z\rho\hat\sigma_i^z|R\rangle
  = m_*(-m_*)\rho_{PR} = -m_*^2\rho_{PR},
\end{equation}
and
\begin{equation}
  -\tfrac{1}{2}\bigl(
    \langle P|\hat\sigma_i^{z\,2}|P\rangle
   +\langle R|\hat\sigma_i^{z\,2}|R\rangle\bigr)\rho_{PR}
  = -\tfrac{1}{2}(1+1)\rho_{PR} = -\rho_{PR}.
\end{equation}
The combined decay rate per spin is
$(-m_*^2-1)\rho_{PR}=-(1+m_*^2)\rho_{PR}$,
and summing over all $N$ spins:
\begin{equation}
  \dot\rho_{PR} = -N\gamma_\phi(1+m_*^2)\,\rho_{PR},
  \qquad
  \Rightarrow\quad \frac{1}{T_2^{\rm local}} = N\gamma_\phi(1+m_*^2).
\end{equation}
For $m_*\to1$ (deep ordered phase) this correctly reduces to
$2N\gamma_\phi$, recovering the fully polarised result.

\textbf{Collective bath.}
With $\hat{J}_z\Pket\approx\frac{Nm_*}{2}\Pket$ and
$\hat{J}_z\Rket\approx-\frac{Nm_*}{2}\Rket$:
\begin{align}
  \langle P|\hat{J}_z\rho\hat{J}_z|R\rangle
    &\approx-\frac{N^2m_*^2}{4}\rho_{PR},\\
  -\tfrac12\langle P|(\hat{J}_z^2\rho+\rho\hat{J}_z^2)|R\rangle
    &\approx-\frac{N^2m_*^2}{4}\rho_{PR}.
\end{align}
Combining: $\dot\rho_{PR}
=-\frac{N^2m_*^2\gamma_\phi}{2}\rho_{PR}$,
confirming Eq.~\eqref{eq:n2_rate}.

\section{Python Code for Five-Level LGI Verification}
\label{app:code}

The following code performs exact diagonalisation of the LMG Hamiltonian,
constructs the Liouvillian for collective dephasing, and computes $K_3$.
All self-tests must pass before results are trusted.
The code directly reproduces all values in Fig.~\ref{fig:multilevel}
and the threshold Eq.~\eqref{eq:gamma_thresh}.
Additional values quoted in the text --- the exact susceptibility
$\chi_{\rm eff}^{\rm exact}\approx1509\,J^{-1}$, ground-state $m{=}0$
population weight ($0.2374\%$), and gap ratio $\DeltaE_{21}/\DeltaE_{10}\approx10.4$
--- are obtained from separate exact-diagonalisation routines; complete
scripts will be deposited to the public repository (Zenodo/GitHub) upon acceptance.
The level-convergence thresholds for $n{=}2,3,4$ are reproducible from the
listing below by setting \texttt{nl=2,3,4} respectively; the values
$0.515,\,0.289,\,0.305$~s$^{-1}$ have been independently verified by
explicit Liouvillian construction using the provided code.
\textbf{Requirements:} Python $\geq3.8$, NumPy $\geq1.20$, SciPy $\geq1.7$;
no additional dependencies beyond the standard scientific Python stack.
The complete code will be deposited to a public archive (Zenodo/GitHub)
upon acceptance; the self-contained listing below is fully executable.

\begin{lstlisting}[
  caption={Exact diagonalisation and five-level LGI calculation.
  All $\gamma_\phi$ in physical units (s$^{-1}$).
  Self-tests, explicit C$_{12}$, C$_{23}$, C$_{13}$ true sequential protocol check, and Goldilocks N-scan included.},
  label=code:lgi]
import numpy as np
from scipy.linalg import eigh, expm
from scipy.optimize import root_scalar

# Requirements: Python >= 3.8, NumPy >= 1.20, SciPy >= 1.7
# No additional dependencies. Run from command line: python lgi_lmg.py

J_PHYS = 37195.4   # rad/s
DELTA_E_P = 1310.0 # rad/s (benchmark from N=370 spins -> 371-dim symmetric Dicke subspace)

def build_lmg(N, Gamma, J=1.0):
    Jt = N/2.0
    m = np.arange(-Jt, Jt+1)
    H = np.diag(-2.0*J/N * m**2)
    for i in range(len(m)-1):
        a = np.sqrt(Jt*(Jt+1) - m[i]*(m[i]+1))
        H[i,i+1] = H[i+1,i] = -Gamma*a
    return H, m

def liouvillian(H, Jz, g):
    d = H.shape[0]; I = np.eye(d, dtype=complex)
    Lc = -1j*(np.kron(I, H.astype(complex)) - np.kron(H.T.astype(complex), I))
    LdL = Jz @ Jz
    Ld = g*(np.kron(Jz.astype(complex), Jz.astype(complex))
            - 0.5*np.kron(LdL.astype(complex), I)
            - 0.5*np.kron(I, LdL.T.astype(complex)))
    return Lc + Ld

vec = lambda r: r.T.flatten()
unvec = lambda v,d: v.reshape(d,d).T

def K3(g_J, dt, H, Jz, Q, rho0):
    # Returns K3 = 2*C12 - C13 (stationarity-approximated correlator, C23=C12 assumed).
    # This equals the true chained K3=C12+C23-C13 when Q^2=I exactly.
    # For the physical Q=sgn(Jz), the difference |C23-C12| < 1% (verified in sequential block below).
    d = H.shape[0]
    prop = expm(liouvillian(H, Jz, g_J)*dt)
    post = 0.5*(Q @ rho0 + rho0 @ Q)
    v0 = vec(post)
    C1 = np.trace(Q @ unvec(prop @ v0, d)).real
    C2 = np.trace(Q @ unvec(prop @ prop @ v0, d)).real
    return 2*C1 - C2

# ---- self-tests ----
def run_tests():
    H2, m2 = build_lmg(2, 0.5)
    ev = np.sort(eigh(H2, eigvals_only=True))
    expected = np.sort([(-1-np.sqrt(5))/2, (-1+np.sqrt(5))/2, -1.0])
    assert np.allclose(ev, expected, atol=1e-10), "FAIL: N=2 eigenvalues"
    print("PASS: N=2 eigenvalues")
    H0 = np.zeros((3,3)); Jz3 = np.diag([-1.,0.,1.])
    L = liouvillian(H0, Jz3, 0.1)
    ri = np.zeros((3,3),dtype=complex); ri[0,2] = 1.0
    rf = unvec(expm(L*2.0) @ vec(ri), 3)
    assert np.isclose(rf[0,2].real, np.exp(-0.4), atol=1e-8), "FAIL: decay"
    print("PASS: dephasing decay")
    rd = np.diag([0.2,0.3,0.5])
    rt = unvec(expm(L*2.0) @ vec(rd), 3)
    assert np.isclose(np.trace(rt).real, 1.0, atol=1e-10), "FAIL: trace"
    print("PASS: trace preservation\n")

run_tests()

# ---- main calculation ----
N, Gamma = 370, 0.95; nl = 5
H_f, mv = build_lmg(N, Gamma)
ev, evec = eigh(H_f)
E = ev[:nl]; dE = E[1]-E[0]
print(f"DeltaE = {dE*J_PHYS:.1f} rad/s (exact: {DELTA_E_P})")
Qd = np.array([np.sign(m) if abs(m) > 0.5 else 0. for m in mv])
Q01_sq = np.vdot(evec[:,0], Qd*evec[:,1]).real**2
print(f"Q01^2 = {Q01_sq:.4f}")
ps =[evec[:,i] for i in range(nl)]
Jzl = np.array([[np.vdot(ps[i],mv*ps[j]) for j in range(nl)] for i in range(nl)]).real
Ql = np.array([[np.vdot(ps[i],Qd*ps[j]) for j in range(nl)] for i in range(nl)]).real
Hl = np.diag(E)
r0 = np.zeros((nl,nl),dtype=complex); r0[0,0]=1.0
dt = np.pi/(3*dE)
gp = np.array([0.005,0.01,0.02,0.05,0.1,0.2,0.3,0.5,1.0,2.0])
gJ = gp/J_PHYS
K3v = np.array([K3(g,dt,Hl,Jzl,Ql,r0) for g in gJ])

def K3_sequential(g_J, dt, H, Jz, Q, rho0):
    """True sequential K3 = C12 + C23 - C13 without stationarity assumption."""
    d = H.shape[0]
    L = liouvillian(H, Jz, g_J)
    prop = expm(L*dt)
    prop2 = expm(L*(2*dt))
    inst = 0.5*(Q @ rho0 + rho0 @ Q)
    v_inst = vec(inst)
    C12 = np.trace(Q @ unvec(prop  @ v_inst, d)).real
    C13 = np.trace(Q @ unvec(prop2 @ v_inst, d)).real
    rho_dt = unvec(prop @ vec(rho0), d)
    inst2 = 0.5*(Q @ rho_dt + rho_dt @ Q)
    C23 = np.trace(Q @ unvec(prop @ vec(inst2), d)).real
    return C12 + C23 - C13

K3v_seq = np.array([K3_sequential(g,dt,Hl,Jzl,Ql,r0) for g in gJ])
print("\ngamma (s^-1) | K3 (5-level, stationary) | K3 (sequential) | diff")
for g,k,ks in zip(gp,K3v,K3v_seq):
    print(f" {g:.3f} | {k:.4f}                   | {ks:.4f}           | {ks-k:+.4f}")
# Note: the stationarity approximation (C23=C12) holds to <1% within the LGI-violating
# regime (gamma <= 0.289 s^-1, where K3>1). At higher dephasing the sequential and
# stationary values diverge, but K3<1 there regardless so the violation conclusion is unaffected.

# --- EXACT ROOT FINDING ---
def K3_continuous(g_phys):
    return K3(g_phys / J_PHYS, dt, Hl, Jzl, Ql, r0) - 1.0
res = root_scalar(K3_continuous, bracket=[0.2, 0.5], method='brentq')
g_thr = res.root
print(f"\nExact 5-level threshold (root-finding): {g_thr:.4f} s^-1")

print(f"K3 at BEC target (0.05 s^-1): {K3v[3]:.4f}")

# ---- Physical T2 from energy-eigenstate matrix elements ----
Jz2_full_diag = mv**2
val0_exact = np.vdot(evec[:,0], Jz2_full_diag * evec[:,0]).real
val1_exact = np.vdot(evec[:,1], Jz2_full_diag * evec[:,1]).real
Gamma_01_exact = 0.5 * (val0_exact + val1_exact)
Jz2_lev = Jzl @ Jzl
Gamma_01_trunc = 0.5*(Jz2_lev[0,0] + Jz2_lev[1,1])
print(f"\nExact full-space: <E0|Jz^2|E0> = {val0_exact:.1f}, <E1|Jz^2|E1> = {val1_exact:.1f}")
print(f"Truncated 5-level (PJzP)^2_00 = {Jz2_lev[0,0]:.1f}, (PJzP)^2_11 = {Jz2_lev[1,1]:.1f}  [reference only; NOT used in T2 formula]")

T2_thr = 1000.0 / (g_thr * Gamma_01_exact)
T2_bec = 1000.0 / (0.05 * Gamma_01_exact)
print(f"T2_phys at threshold = {T2_thr:.3f} ms = {T2_thr/1000*DELTA_E_P:.3f} hbar/DeltaE")
print(f"T2_phys at BEC target (0.05 s^-1) = {T2_bec:.3f} ms")

# ---- 10-level convergence check ----
print("\n=== 10-level convergence check ===")
nl10 = 10
ps10 = [evec[:,i] for i in range(nl10)]
Jzl10 = np.array([[np.vdot(ps10[i],mv*ps10[j]) for j in range(nl10)] for i in range(nl10)]).real
Ql10 = np.array([[np.vdot(ps10[i],Qd*ps10[j]) for j in range(nl10)] for i in range(nl10)]).real
Hl10 = np.diag(ev[:nl10])
r010 = np.zeros((nl10,nl10),dtype=complex); r010[0,0]=1.0
K3v10 = np.array([K3(g,dt,Hl10,Jzl10,Ql10,r010) for g in gJ])
print("gamma (s^-1) | K3 (5-level) | K3 (10-level) | diff")
for g,k5,k10 in zip(gp,K3v,K3v10):
    print(f" {g:.3f} | {k5:.4f} | {k10:.4f} | {k10-k5:+.4f}")

def K3_continuous10(g_phys):
    return K3(g_phys / J_PHYS, dt, Hl10, Jzl10, Ql10, r010) - 1.0
res10 = root_scalar(K3_continuous10, bracket=[0.2, 0.5], method='brentq')
g_thr10 = res10.root
print(f"\nExact 10-level threshold: {g_thr10:.4f} s^-1")
print(f"Threshold shift n=5 -> n=10: {100*(g_thr10-g_thr)/g_thr:+.1f}% (exact); using rounded 0.289/0.286: ~1% (convergence confirmed)")

# === Explicit C12, C23, C13 with TRUE SEQUENTIAL PROTOCOL ===
# To properly test LGI, we evolve undisturbed to t1 (dt), measure (disturbing the state),
# and then evolve to t2 (2*dt) to find C23.
g_test = gJ[3]

# 1. C12 and C13: Apply measurement instrument at t=0, evolve forward
inst1 = 0.5 * (Ql @ r0 + r0 @ Ql)
v_inst1 = vec(inst1)

rho_dt = unvec(expm(liouvillian(Hl, Jzl, g_test) * dt) @ v_inst1, nl)
C12 = np.trace(Ql @ rho_dt).real

rho_2dt = unvec(expm(liouvillian(Hl, Jzl, g_test) * (2*dt)) @ v_inst1, nl)
C13 = np.trace(Ql @ rho_2dt).real

# 2. C23: Evolve UNDISTURBED to dt, apply measurement instrument, evolve another dt
rho_dt_undisturbed = unvec(expm(liouvillian(Hl, Jzl, g_test) * dt) @ vec(r0), nl)
inst2 = 0.5 * (Ql @ rho_dt_undisturbed + rho_dt_undisturbed @ Ql) # Measurement at dt
v_inst2 = vec(inst2)

rho_2dt_from_dt = unvec(expm(liouvillian(Hl, Jzl, g_test) * dt) @ v_inst2, nl)
C23 = np.trace(Ql @ rho_2dt_from_dt).real

K3_chained = C12 + C23 - C13
K3_simplified = 2 * C12 - C13

print("\n=== Explicit C12, C23, C13 (True Sequential Protocol) ===")
print(f"C12             = {C12:.6f}")
print(f"C23             = {C23:.6f}")
print(f"C13             = {C13:.6f}")
print(f"K3 (chained)    = {K3_chained:.6f}")
print(f"K3 (simplified) = {K3_simplified:.6f}")
print(f"|C23 - C12|     = {abs(C23 - C12):.4f} (Difference due to measurement back-action)")

# ---- Goldilocks N-scan ----
def goldilocks_scan():
    print("\n=== Goldilocks N-scan (Gamma/J = 0.95) ===")
    T_nK = 10.0
    kB_T_phys = 131.0 * T_nK  # ~1310 rad/s at 10 nK
    print(f"Target kB*T = {kB_T_phys:.1f} rad/s")
    for N_test in range(200, 401, 50):
        H_test, _ = build_lmg(N_test, 0.95)
        ev_test = eigh(H_test, eigvals_only=True, subset_by_index=[0,1])
        dE_test = (ev_test[1] - ev_test[0]) * J_PHYS
        print(f"N = {N_test}: DeltaE = {dE_test:.1f} rad/s")
goldilocks_scan()
\end{lstlisting}

\end{document}